\documentclass[aps,superscriptaddress,twocolumn,twoside,floatfix,pra,a4paper]{revtex4-2}
\usepackage{times}
\usepackage{epsfig}
\usepackage{amsfonts}
\usepackage{amsmath}
\usepackage{amssymb,amsthm}
\usepackage{color}
\usepackage{multirow}
\usepackage{braket}
\usepackage{bbm}
\usepackage{latexsym}
\usepackage{amsfonts}
\usepackage{mathrsfs}
\usepackage{natbib}
\usepackage{verbatim}
\usepackage{gensymb}
\usepackage{caption}
\usepackage{subcaption}
\usepackage{subcaption}
\usepackage{graphicx}
\allowdisplaybreaks

\usepackage[colorlinks=true,linkcolor=blue,citecolor=magenta,urlcolor=blue]{hyperref}
\allowdisplaybreaks

\begin{document}

\title{Testing Whether Gravity Acts as a Quantum Entity When Measured}
\author{Farhan Hanif}
\email{farhan.hanif.17@ucl.ac.uk}
\thanks{These two authors contributed equally to this work.}
\affiliation{Department of Physics and Astronomy, University College London, Gower Street, London WC1E 6BT, England, United Kingdom}

\author{Debarshi Das}
\email{debarshi.das@ucl.ac.uk}
\thanks{These two authors contributed equally to this work.}
\affiliation{Department of Physics and Astronomy, University College London, Gower Street, London WC1E 6BT, England, United Kingdom}

\author{Jonathan Halliwell}
\affiliation{Blackett Laboratory, Imperial College, London SW7 2BZ, England, United Kingdom}

\author{Dipankar Home}
\affiliation{Center for Astroparticle Physics and Space Science (CAPSS), Bose Institute, Kolkata 700 091, India}

\author{Anupam Mazumdar}
\affiliation{Van Swinderen Institute, University of Groningen, 9747 AG Groningen, The Netherlands}

\author{Hendrik Ulbricht}
\affiliation{School of Physics and Astronomy, University of Southampton, Southampton SO17 1BJ, England, United Kingdom}

\author{Sougato Bose}
\affiliation{Department of Physics and Astronomy, University College London, Gower Street, London WC1E 6BT, England, United Kingdom}


\begin{abstract}
A defining signature of classical systems is 
``in principle measurability" without disturbance: a feature manifestly violated by quantum systems.  We describe a multi-interferometer experimental setup that can, in principle, reveal the nonclassicality of a spatial superposition-sourced gravitational field if an irreducible disturbance is caused by a measurement of gravity. While one interferometer sources the field, the others are used to measure the gravitational field created by the superposition. 
This requires neither any specific form of nonclassical gravity, nor the generation of entanglement between any relevant degrees of freedom at any stage, thus distinguishing it from the experiments proposed so far. This test, when added to the recent entanglement-witness based proposals, enlarges the domain of quantum postulates being tested for gravity. 
Moreover, the proposed test yields a signature of quantum measurement induced disturbance for any finite rate of decoherence, and is device independent.
\end{abstract}
\maketitle

{\em Introduction:} As far as empirical evidence is concerned, nature is described accurately as a hybrid of quantum field theories (all matter and three of the forces) and a classical theory of gravity (general relativity). However, matter sources gravity, and thereby an unresolved age old question is whether the gravitational field of a mass in a spatial quantum superposition is quantum or classical \cite{feynman1957the,penrose1996gravity,diosi1989models,oppenheim2022constraints,10.3389/fphy.2022.891977,bose2023massive}. ``Ruling out'' gravity as a classical field or curvature by creating large enough masses in such quantum superpositions, although challenging
\cite{bose1999scheme,scala2013matter,wan2016free,pedernales2020motional,marshman2022constructing,margalit2020realization,zhou2022catapulting,zhou2022mass,bose2023massive}, is potentially less demanding than detecting quantum corrections to gravitational interactions \cite{donoghue1995introduction} or on-shell gravitons \cite{parikh2021signatures,parikh2021quantum,Tobar2023detecting,Carney2024graviton}. In this respect, a major progress has been made recently, with the proposal to entangle two masses in quantum superpositions through their gravitational interaction \cite{bose2016matter,bose2017spin,marletto2017gravitationally}. Although the gravitational interaction between the masses is, to any degree of near-term testability, purely Newtonian, it can be argued that the generation of this entanglement between the masses necessitates a quantum superposition of geometries \cite{christodoulou2019possibility}. 
Several persuasive arguments have been put forward linking this experiment with the nonclassicality of gravity  \cite{marshman2020locality,bose2022mechanism,belenchia2018quantum,carney2022newton,danielson2022gravitationally,galley2022no,christodoulou2022locally} and several variants have been proposed \cite{qvarfort2020mesoscopic,krisnanda2022quantum,carney2021using,Biswas2022Gravitational,etezad2023paradox, Kent2022bell}.

\begin{figure*}[htbp!]
\centering
\includegraphics[height=6.3cm,width=13.1cm]{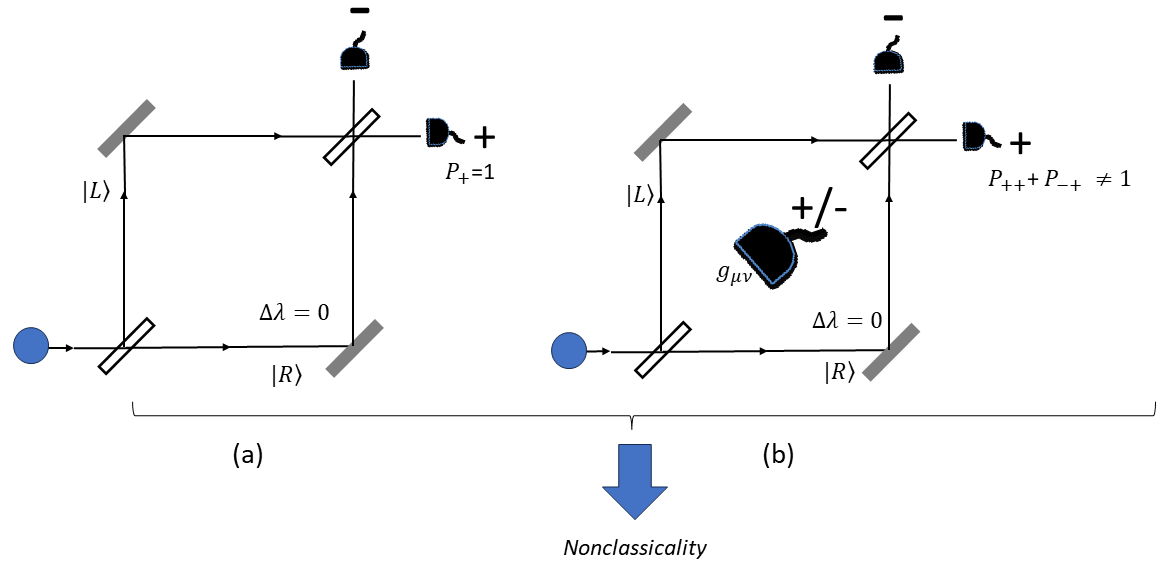}
\caption{A  source mass is prepared in a superposition of states $\ket{L}$ and $\ket{R}$ by subjecting it through an ideal Mach-Zehnder interferometer, while ensuring no interferometric phase difference between the arms ($\Delta \lambda = 0$). 
	(a) Given that no intermediate measurement is performed, the final detector outcome is certain to be $+$: $P_{+} = 1$. 
	(b) An intermediate measurement of the gravitational field of the  source  mass is performed by a suitable detector (Schematically shown as the large detector measuring the metric $g_{\mu \nu}$).   This  measurement has two outcomes ($\pm$). {\em If}, after this intermediate detection, the final outcome probability (averaged over outcomes of the intermediate measurement) differs from unity, it implies that gravity is nonclassical.} 
\label{fig:Fig-1-Macrorealism}
\end{figure*}

The principal obstacle of the above proposal \cite{bose2016matter,bose2017spin,marletto2017gravitationally} is decoherence.  If the decoherence rate $\Gamma >d\Delta \phi/dt$, where $d\Delta \phi/dt$ is the rate of growth of the phase responsible for gravity induced entanglement, then no entanglement is produced between the masses \cite{schut2022deco,van2020quantum,Rijavec_2021} (verifiable using the Peres-Horodecki criterion \cite{Peres1996sep,HORODECKI19961}).   Moreover, witnessing entanglement requires trusted measurement devices. Although one may use device-independent detection of entanglement through the Bell test \cite{Kent2022bell}, that demands an even lower decoherence rate \cite{supp}, as well as closing all loopholes, which is challenging. Thus the key question is whether some {\em other} nonclassical aspect of gravity can be observed in the $\Gamma >d\Delta \phi/dt$ regime, which will be detectable much {\em earlier} in experiments. Notably, a coherence $\sim e^{-\Gamma t}$ is always present in any spatial superposition of a mass evolved for a time $t$. Can that be exploited to  observe some nonclassicality of gravity?  Motivated thus, here we propose to test a different nonclassical aspect of gravity, which is, at the same time, a device-independent test, and works for any finite decoherence rate.  While entanglement witnessing \cite{bose2017spin,marletto2017gravitationally} tests the validity of  quantum superposition principle for gravity, our present proposal can test {\em whether} a measurement of gravity generically causes disturbance (an irreducible feature of quantum measurement).

As quantum mechanics is not defined by the superposition principle alone, but also requires the unitarity of evolution and the measurement postulate \cite{Nielsen_Chuang_2010,PhysRevLett.126.110402}, witnessing entanglement in the earlier proposal \cite{bose2017spin,marletto2017gravitationally} will imply that gravity is described either by quantum mechanics, or by a (unknown) nonclassical theory that obeys superposition principle. To know whether gravity is indeed quantum, we need to test other quantum mechanical postulates for gravity. This is a gap in the literature that we hereby fill by proposing to test a specific aspect of the quantum measurement postulate, namely, quantum measurement-induced disturbance. Adding this test  to the entanglement-witness based test \cite{bose2017spin,marletto2017gravitationally} will take us towards a more {\em complete} demonstration of gravity as a quantum entity.


An ideal measurement on a classical field  should not, {\em in principle}, alter the state of any system (other than, obviously, the state of the probe which registers the field) \cite{steane}. In fact, that should be taken as a crucial part of the definition of any classical field, followed from our everyday notion of classicality \cite{PhysRevLett.54.857}.
This leads to  the  testable ``nondisturbance condition" (NDC)  \cite{NatComm.16,PhysRevA.87.052115,PhysRevA.92.032101}: The act of performing an intermediate measurement should not influence the outcome statistics of a subsequent measurement. Observing a discrepancy between intermediately measured and intermediately unmeasured statistics would thus be  a signature of nonclassicality.  In practice, a clumsy measurement on a classical field can cause disturbance (classical disturbance). Crucially, this disturbance is not an inherent part of classical physics--one can arbitrarily reduce it by performing the measurement appropriately. On the other hand, the quantum measurement-induced disturbance is an intrinsic part of quantum theory, which {\em cannot} be eliminated by any means. This feature is central  to our proposal to show the irreducible nonclassicality of gravity.   

{\em Schematics:}  We first present the general idea as a schematic. A source mass described by quantum mechanics, but large
enough to produce a detectable gravitational field at a proximal detector, is made to undergo an interferometry with equal
amplitudes in the arms (labeled by quantum states $|L\rangle$
and $|R\rangle$). The outputs at the
end of the interferometry (which could be direct electromagnetic detection of the source mass) are labeled $+$ and $-$, while
the relative phase $\Delta \lambda$ between the arms is ensured to be $0$.
This setting [Fig.\ref{fig:Fig-1-Macrorealism}(a)] is then compared with another
setting [Fig.\ref{fig:Fig-1-Macrorealism}(b)], where an intermediate gravitational field detector is placed during the interferometry. In practice, the most
sensitive such detector will be similar mass (masses) undergoing interferometry (interferometries). {\em It is crucial to ensure that the detector performs an intermediate measurement (midway during the interferometry) of the gravitational field of the source mass rather than the position of the source mass itself by other means (i.e., via electromagnetic channels,  or scattered photons).} Without considering any specificity of the information obtained through the measurement, we assume that this measurement gives one bit of information about the gravitational field with outcomes depicted by $+$ and $-$. Subsequently, a detection of the source mass is also made in the $+$ and $-$ outputs of the interferometer. If a ``hybrid model'' is used with quantum matter, but classical gravity, then, by definition (of classicality), the measurement of gravity by the intermediate detector cannot cause any change in the final probabilities, i.e., 
\begin{equation}
P_{+} (\text{no intermediate meas})-P_{+} (\text{after intermediate meas})=0,
\label{nsit-basic}
\end{equation} 
where $P_{+} (\text{after intermediate meas})=P_{+,+}+P_{-,+}$ (here $P_{a,b}$ is the joint probability of getting the outcomes $a$, $b$ in the intermediate and the final measurements respectively).
Equation (\ref{nsit-basic}) is the NDC to be satisfied by gravity as a classical entity. Any violation of this NDC implies that gravity is nonclassical. Here,  we must ensure that  $\Delta \lambda =0$ is still maintained while going from the case of Fig.\ref{fig:Fig-1-Macrorealism}(a) to Fig.\ref{fig:Fig-1-Macrorealism}(b) even though an extra intermediate detector is coupled, as otherwise the probability of $P_{+}$ can simply change due to an interferometric phase difference rather than due to the measurement.

Any NDC violation in our experiment will rule out hybrid models (classical gravitational field sourced by quantum matter) for which the gravitational field can, by definition, be measured without disturbance. Examples of hybrid models \cite{oppenheim2022constraints,Moller1962Les,ROSENFELD1963353,kafri2014classical,PhysRevLett.81.2846,penrose1996gravity} satisfying NDC can be found in \cite{supp}.  Here we emphasize the necessity of both parts of the experiment. Figure \ref{fig:Fig-1-Macrorealism}(a) alone reveals nothing about the {\em form} of gravity sourced by the source mass as no gravitational field is measured at any stage. On the other hand, Fig.\ref{fig:Fig-1-Macrorealism}(b) alone does not tell whether the source mass superposition has already been affected even before the measurement (e.g., as in a spontaneous collapse model \cite{penrose1996gravity}). Thus any proposal involving Fig.\ref{fig:Fig-1-Macrorealism}(b) alone, {\em without} comparing to Fig.\ref{fig:Fig-1-Macrorealism}(a) (e.g. \cite{PhysRevLett.47.979}) is insufficient on its own to reveal nonclassicality of gravity.

\begin{figure}[t!]
\includegraphics[height=8cm,width=8cm]{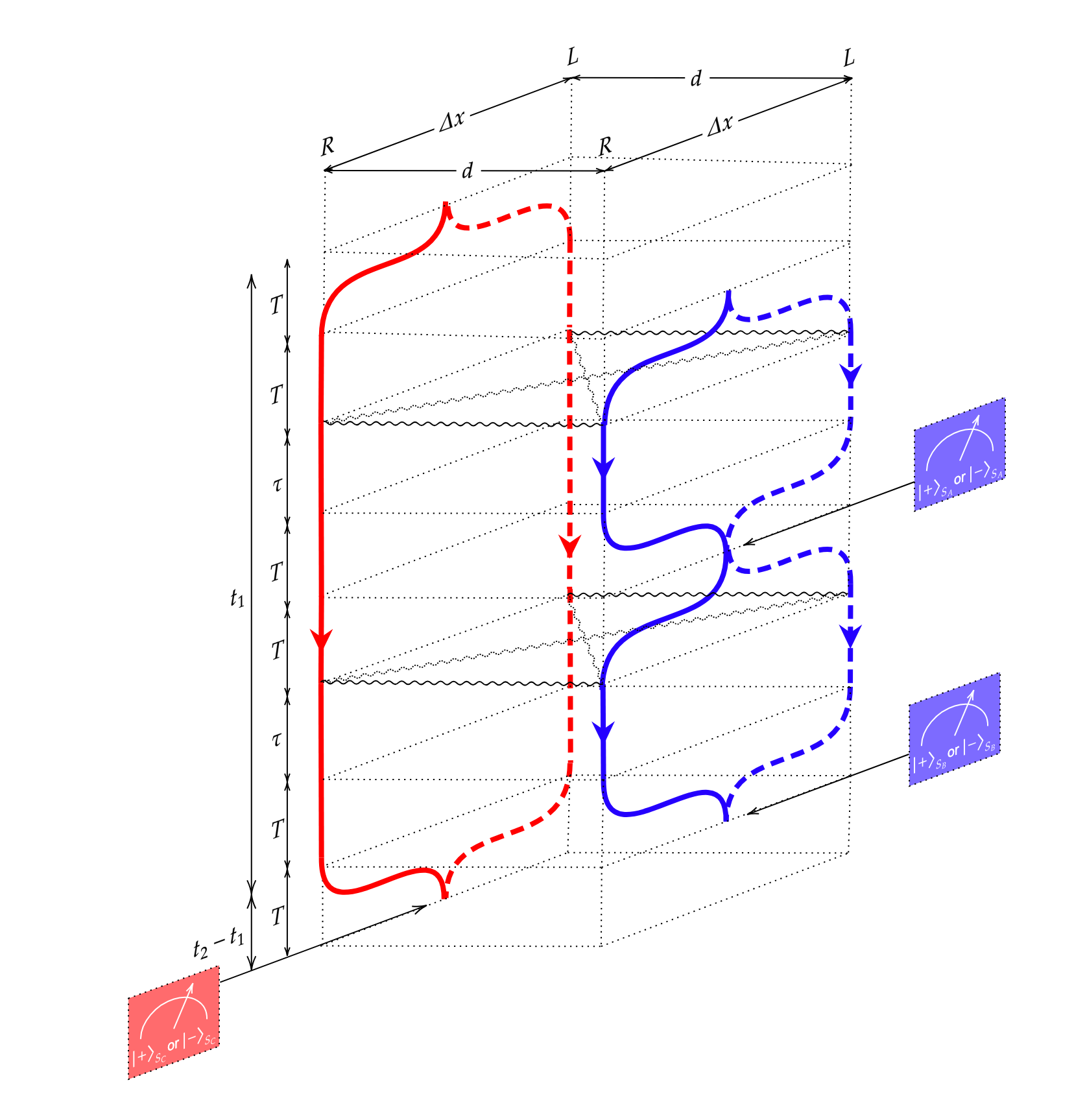} 
\caption{The gravitational field generated by the interferometric source mass (red) is measured sequentially by a pair of massive interferometric probes (blue), where  the gravitational interactions are indicated by wavy lines. 
	Finally, the source mass superposition is closed and a measurement is performed on the embedded spin of the source mass. 
}
\label{twoprobe}
\end{figure}

\emph{Interferometric setup:} We consider a specific arrangement in which the source mass $M$ with an embedded spin undergoes a spin dependent spatial interferometry (also called a Stern-Gerlach interferometry \cite{margalit2020realization}). This replaces the Mach-Zehnder interferometer depicted in Fig.\ref{fig:Fig-1-Macrorealism}. The unmeasured case [corresponding to Fig.\ref{fig:Fig-1-Macrorealism}(a)] of the experiment is performed only with this mass. The intermediate detector for measuring the gravitational field of the source mass [corresponding to Fig.\ref{fig:Fig-1-Macrorealism}(b)] is realised by two successive probe interferometers, each with mass $m$ and an embedded spin, arranged in a geometrically parallel configuration with respect to the source interferometer at some distance $d$ away. The spatial superposition of the source mass is then closed and a projective measurement is performed on its embedded spin. The protocol is depicted in Fig.\ref{twoprobe}. We finally compare the statistics of the final spin measurement with and without the intermediate gravitational field measurements to test the NDC.

All masses are prepared, held in spatial superposition (mechanism to create such superposition can be found in \cite{bose1999scheme,scala2013matter,wan2016free,pedernales2020motional,marshman2022constructing,margalit2020realization,zhou2022catapulting,zhou2022mass,bose2023massive}), and recombined for completing interferometry through specific means, such as spin motion coupling. 
In what follows, let $M_i$ and $S_i$ denote the mass and embedded spin degrees of freedom of a given mass indexed by $i$ according to whether one of the two probe systems ($i=A,B$ in sequence) or the source system ($i=C$) is referenced.

The initial state of the source mass with its embedded spin at $t=0$ is given by,
\begin{equation}
\ket{\psi(t=0)} = \ket{\zeta}_{M_C}\otimes\dfrac{1}{\sqrt 2}(\ket{\uparrow}_{S_C}+\ket{\downarrow}_{S_C})\label{Init}, \nonumber 
\end{equation}
where $\ket{\zeta}_{M_C}$ is the initial localized state of the source mass at the center of
the axis of the source interferometer. Over a time $T$, the source mass is prepared in spatial superposition via the unitary evolution:
\begin{align}
\ket{\zeta}_{M_C}\otimes\ket{\uparrow}_{S_C} \to \ket{L \uparrow}_{C}, \, \, 
\ket{\zeta}_{M_C}\otimes\ket{\downarrow}_{S_C} \to \ket{R \downarrow}_{C}. \label{evo}
\end{align}
In the above, the states $|L\uparrow\rangle_C$ and $|R\downarrow\rangle_C$ are separated by 
a distance $\Delta x (t)$, which grows from $0$ at $t=0$ to the maximum at $t=T$ with $\Delta x (T) = \Delta x$.
The first probe mass $M_A$ (of mass $m$) with embedded spin $S_A$ is then introduced and subjected to the evolution (\ref{evo}) with the subscript `$C$' being replaced by `$A$' over another time interval $T$. 

With both superpositions fully prepared, the source and the probe now interact exclusively through gravity in a static geometrical arrangement for a time $\tau$ before the spatial  superposition of the probe is closed over a time $T$ \cite{bose2017spin,margalit2020realization}. Thus the total interaction time interval is $2 T + \tau$.  At this stage, the joint state of the source and probe is given by
\begin{align}
\ket{\psi}_{C,A} =&\dfrac{1}{\sqrt{2}}\biggl(\sqrt{1+ \cos \Delta \phi} \ket{\Psi_+}_C\ket{+}_{S_A} \nonumber \\
&+ \sqrt{1- \cos \Delta \phi}\ket{\Psi_-}_C\ket{-}_{S_A}\biggr) \ket{\zeta}_{M_A},
\label{source-probe-ent}
\end{align}
with
\begin{align}
&\ket{\Psi_{\pm}}_{C} =\dfrac{\left(1 \pm e^{i\Delta \phi} \right) \ket{L\uparrow}_{C}+\left(e^{i\Delta \phi}\pm 1\right)\ket{R\downarrow}_{C}}{2\sqrt{1\pm\cos \Delta \phi}} \nonumber \\
&\ket{\pm}_{S_A} = \dfrac{\ket{\uparrow}_{S_A} \pm \ket{\downarrow}_{S_A}}{\sqrt{2}},
\label{source-probe-ent-2}
\end{align}
where $\Delta \phi = \Delta \phi_{\tau} + 2\Delta \phi_T$ is a function of the relative phases accumulated between the different arms of the source and each of the probe interferometers over their total interaction time duration $2T + \tau$. Of its constituent parts, $\Delta \phi_T$ is the relative phase accumulated during the opening or the closing of the spatial  superposition of each probe, with its expression being somewhat elaborate (given in \cite{supp}), while $\Delta \phi_\tau$ is associated with the relative phase development for the duration $\tau$ when the spatial superpositions of source and each probe are held in a static geometrical arrangement and is given by,
\begin{equation}
\Delta \phi_{\tau} =\dfrac{GMm\tau}{\hbar\sqrt{d^2 + (\Delta x)^2}} - \dfrac{GMm\tau}{\hbar d}.
\end{equation} 

Note that the probe mass is not affected by contact (or otherwise electromagnetically) with the source mass, but only being affected at a distance by the source's gravity (i.e., through the metric $g_{00}$, which is completely determined by the source mass). After closure of the interferometry of the probe, its spin state is decoupled from its spatial state which enables accessing the information about the relative phases accumulated between $|L\uparrow\rangle_A$ and $|R\downarrow\rangle_A$ due to gravitational interaction between the source and the probe. Accordingly, 
a projective measurement of the probe spin is now performed in the $\ket{\pm}_{S_A}$ basis. This projection results in a POVM on the source system (mass and its associated field). Since only the gravitational field of the source is in contact with the probe, we can say that this POVM is essentially a measurement of gravity.

The first probe  is then  discarded, and a new probe  is introduced. As before, the new probe now interacts with the source system via the gravitational field for a further time $2T+\tau$ in an identical fashion before a projective measurement in the $\ket{\pm}_{S_B} = (\ket{\uparrow}_{S_B}\pm \ket{\downarrow}_{S_B})/\sqrt{2}$ basis is performed on the spin degree of freedom of the second probe at  $t= t_1 = 5 T + 2 \tau$. As argued earlier, this is also a measurement of the source's gravity. 
The second probe is then also discarded. Over a time $T$, the spatial superposition of the source interferometer is now closed via the reversal of the unitary evolution \eqref{evo}. 

A final projective measurement of the source spin is then performed in the $\ket{\pm}_{S_C} = (\ket{\uparrow}_{S_C}\pm \ket{\downarrow}_{S_C})/\sqrt{2}$ basis at $t=t_2$ where $t_2-t_1=T$. 
This measurement  yields the following unnormalized states of the source conditioned on the outcomes of the three measurements (for details, see \cite{supp}):
\begin{align}    
&\ket{\psi_{a,b,c}}= \dfrac{1}{8}\biggr[\biggl(1+ae^{i\Delta \phi}\biggr)\biggl(1+be^{i\Delta \phi}\biggr)\nonumber\\
& \hspace{0.2cm}+ c e^{2i\Delta \phi}\biggl(1+a e^{-i\Delta \phi}\biggr)\biggl(1+ b e^{-i\Delta \phi}\biggr) \biggr]\ket{\zeta}_{M_C}\ket{c}_{S_C},\label{final1} 
\nonumber
\end{align}
where $a,b,c\in\{+,-\}$ denote the outcomes of the first and second probe measurements followed by the  final measurement on the source spin respectively. From the norms of these states, the joint probabilities $P_{a,b,c}$ are obtained. 

Let us now consider the same scenario as described above, except that the probes are not introduced, and thus no intermediate measurement takes place prior to the final measurement on the source spin at $t=t_2$. 
In this case, the probabilities of the final measurement outcomes are $P_{+}=1$, $P_{-}=0$.

Thus the violation of the NDC is given by \cite{supp},
\begin{equation}
V(\pm) = P_{\pm} - \sum_{a,b\in\{\pm\}}P_{a,b,\pm} = \pm\dfrac{1}{2}\sin^2\Delta \phi\label{Vpm1}.
\end{equation}
This NDC violation implies that measurement of gravity causes disturbance. Notably, NDC violation persists (although suppressed) for any finite rate of decoherence \cite{supp}.

This is a device-independent test of nonclassicality in the sense that the intermediate and the final measurements need not to be trusted. We only need to ensure that the intermediate measurements are on the source's gravitational field.

While the calculations \cite{supp} are carried out under the application of an instantaneous, manifestly nonlocal Newtonian field, this is merely a {\em calculational tool} that yields outcomes consistent with a relativistic description \cite{marshman2020locality,bose2022mechanism,christodoulou2022locally}.



\emph{Is entanglement between the source and the probe necessary?:} Equation (\ref{source-probe-ent}) implies that entanglement is created between the source and the first probe (similarly for the second probe). This is obtained following the usual quantum formalism and is the core of the earlier proposal \cite{bose2017spin,marletto2017gravitationally}. Now, let us consider another hypothetical nonclassical theory of gravity (different from quantum theory), where the gravitational interaction between the source and the probe produces the following separable joint state (following some unknown mechanism),
\begin{align}
&\rho_{C,A} =\dfrac{1}{2}\biggl(\left(1+ \cos \Delta \phi \right) \ket{\Psi_+}_C \bra{\Psi_+}_C \otimes  \ket{+}_{S_A}  \bra{+}_{S_A} + \nonumber \\
&\left(1- \cos \Delta \phi\right)\ket{\Psi_-}_C \bra{\Psi_-}_C \otimes \ket{-}_{S_A}  \bra{-}_{S_A}\biggr) \otimes \ket{\zeta}_{M_A} \bra{\zeta}_{M_A}
\nonumber
\end{align}
In this case, classical correlation created between the source and the probe is sufficient to perform measurement of the source's gravity. Following similar gravitational interaction between the source and the second probe, the same NDC violation (\ref{Vpm1}) is obtained. If gravity obeys such a nonclassical theory, then the previous proposal \cite{bose2017spin,marletto2017gravitationally} fails as no gravity-induced entanglement is generated. However, the present proposal can witness nonclassicality of gravity in such a case. This establishes the independence of the present proposal with respect to the previous one \cite{bose2017spin,marletto2017gravitationally}.

\emph{Why two probes:} Quantum measurements, accompanied by an averaging over the outcomes, essentially cause a {\em dephasing} of the source mass. This is mathematically equivalent to a probabilistic phase flip, with the probability of phase flip growing from $0$ initially to $1/2$ at infinite time (complete dephasing). This is indeed at the core of violating NDC. However, we should prevent any additional deterministic phase (equivalent to $\Delta \lambda \neq 0$) caused by the presence of the probe as it can be interpreted as a classical disturbance due to a common gravitational acceleration experienced by both $|L\rangle_{M_C}$ and $|R\rangle_{M_C}$ of the source mass \cite{colella1975observation}. In our proposal, two separate probe measurements are employed to eliminate this classical disturbance \cite{supp}.

\emph{Parameter regimes:} To exemplify, let us consider the parameter regime with $M, m \sim 10^{-14}$ kg, and closest approach of the masses $d\sim 157$ $\mu$m to ensure that gravity is significantly stronger than the electromagnetic interactions between neutral masses \cite{van2020quantum} such that the intermediate measurements are indeed on the gravitational field. As the  superposed trajectories in each interferometer are fixed through magnetic gradients \cite{margalit2020realization,scala2013matter,wan2016free,pedernales2020motional,marshman2022constructing,zhou2022catapulting,zhou2022mass}, which is much stronger than the gravitational force between the masses, we can safely assume that the gravitational pull on the source due to the probe (a classical disturbance) is negligible. In practice, we must further ensure the following \cite{NatComm.16}: acting as a control experiment, a classical mixture of the two localized states $|L\rangle_{M_C}$ and $|R\rangle_{M_C}$ of the source mass should be prepared instead of a superposition, which is expected to give rise to a classical-like gravitational field. Then the detected NDC violation (which arises solely due to classical disturbance and would give zero in the ideal case) should be ensured to be at least $1$ order of magnitude less than the detected NDC violation obtained by preparing the spatial superposition of the source mass under the same experimental conditions. For negligible decoherence, NDC violation $\gtrsim 0.4$ can be obtained with $\tau,T \sim 1.9-3.2$ s, and $\Delta x \sim 215-479\mu$m (see \cite{supp} for details, including
effects of decoherence). One can reduce $M, m, d$, and/or $\Delta x$ by a few orders of magnitude \cite{schut2023micron} keeping the same violations. It may be easier for experiments to reduce $M, m$, and $\Delta x$ at the price of increasing the number of runs.  As NDC violation effectively amounts to measuring probabilities, we can measure a lower violation of $0.01$ by averaging the results of $>10^4$ experimental runs.   
The requirements on pressures, temperatures and inertial noises to keep the decoherence negligible in the context of the earlier proposal \cite{bose2017spin,van2020quantum,torovs2020relative} are not strictly necessary for the present proposal, as NDC violation persists for any finite decoherence rate. For example, for the typical parameter choice of the earlier proposal \cite{bose2017spin,marletto2017gravitationally}, generation of entanglement requires $\Gamma t <  10^{-2}$ (with $t$ being the total interaction time) \cite{van2020quantum,schut2023micron}, which is equivalent to keeping the vacuum pressure $P < 5 \times 10^{-16}$ Pa. On the other hand, in our proposal $\Gamma t \sim  1$ (equivalently, $P \sim 5 \times 10^{-14}$ Pa) can give substantial violation of the NDC.

One drawback of the present proposal (and also of the previous proposal \cite{bose2016matter,bose2017spin,marletto2017gravitationally}) is that the mass of the experimental apparatus (e.g., the magnets in the Stern-Gerlach interferometers, etc.) is ignored. However, the mass of the apparatus can cause backaction on the interference of the masses due to the equivalence principle \cite{Bronstein2012republication,Torrieri2023equivalence}, which may have an adverse effect on our proposal. Hence, considering this effect \cite{Torrieri2023equivalence} in the context of the present proposal merits further investigation.

\emph{Conclusions:} There is an existing proposal for testing the validity of the quantum superposition principle for gravity via witnessing gravity-induced entanglement \cite{bose2017spin,marletto2017gravitationally}. Here, we have suggested a scheme which will {\em complement} that test by showing that when gravity is measured, there is an irreducible disturbance (a nonclassical feature). As we are summing over the measurement-outcomes for testing NDC, the measurement is equivalent to decoherence, but a decoherence which is controllably triggered only by the act of measurement \cite{text}.  We should point out that  our present work is different from \cite{matsumura2022leggett} where the violation of Leggett-Garg inequalities (a class of inequalities violated by nonclassical theories) is used to infer gravity-induced entanglement. The quantum disturbance due to measurement of gravity is not sought to be tested there. 


The earlier proposal \cite{bose2017spin,marletto2017gravitationally} tests only the final entanglement between the spins of the two masses and does not fully specify the dynamics needed to reach the state. Hence, this earlier proposal cannot verify that the probe can measure the gravitational field of the source causing an irreducible disturbance. This new physical insight will be obtained by realizing the present proposal. If the decoherence rate is too high such that no entanglement is generated between the two masses, then the earlier proposal \cite{bose2017spin,marletto2017gravitationally} fails in the sense that gravity cannot be concluded as a nonclassical communication channel acting between the two masses. In such extreme cases also, the correlation (weaker than entanglement) generated between the source and the probe enables us to perform a measurement of gravity, which inevitably causes disturbance leading to observable violation of NDC. Further, our test enables us to capture nonclassicality of gravity in a landscape of theories which are neither classical (violating NDC), nor fully quantum (fundamentally unable to generate entanglement between two masses). Thus the other test \cite{bose2017spin,marletto2017gravitationally} should complement the present one to proceed towards capturing the full quantumness of gravity.


\begin{acknowledgments}
{\it Acknowledgements:--} We acknowledge fruitful discussions with Lev Vaidman and Adrian Kent. F.H. acknowledges support from the Engineering and
Physical Sciences Research Council (Grant No.
EP/L015242/1). D.D. acknowledges the fruitful discussions with Marko Toro\v{s} and Lorenzo Braccini. D.D. also acknowledges the Royal Society (United Kingdom) for the support through the Newton International Fellowship (NIF$\backslash$R$1\backslash212007$). J.H. acknowledges useful conversations with Clement Mawby. D.H. acknowledges support from NASI Senior Scientist Fellowship and QuEST-DST Project Q-98 of the Government of India. H.U. would like to acknowledge support from EPSRC through Grants No. EP/W007444/1, No. EP/V035975/1 and No.  EP/V000624/1, the Leverhulme Trust (RPG-2022-57), the EU Horizon 2020 FET-Open project TeQ (766900), and the EU EIC Pathfinder project QuCoM (10032223). S.B. would like to acknowledge EPSRC Grant No. EP/X009467/1 and STFC Grant No. ST/W006227/1.
\end{acknowledgments}

\bibliography{ref2} 

\begin{thebibliography}{63}%
\makeatletter
\providecommand \@ifxundefined [1]{%
 \@ifx{#1\undefined}
}%
\providecommand \@ifnum [1]{%
 \ifnum #1\expandafter \@firstoftwo
 \else \expandafter \@secondoftwo
 \fi
}%
\providecommand \@ifx [1]{%
 \ifx #1\expandafter \@firstoftwo
 \else \expandafter \@secondoftwo
 \fi
}%
\providecommand \natexlab [1]{#1}%
\providecommand \enquote  [1]{``#1''}%
\providecommand \bibnamefont  [1]{#1}%
\providecommand \bibfnamefont [1]{#1}%
\providecommand \citenamefont [1]{#1}%
\providecommand \href@noop [0]{\@secondoftwo}%
\providecommand \href [0]{\begingroup \@sanitize@url \@href}%
\providecommand \@href[1]{\@@startlink{#1}\@@href}%
\providecommand \@@href[1]{\endgroup#1\@@endlink}%
\providecommand \@sanitize@url [0]{\catcode `\\12\catcode `\$12\catcode
  `\&12\catcode `\#12\catcode `\^12\catcode `\_12\catcode `\%12\relax}%
\providecommand \@@startlink[1]{}%
\providecommand \@@endlink[0]{}%
\providecommand \url  [0]{\begingroup\@sanitize@url \@url }%
\providecommand \@url [1]{\endgroup\@href {#1}{\urlprefix }}%
\providecommand \urlprefix  [0]{URL }%
\providecommand \Eprint [0]{\href }%
\providecommand \doibase [0]{https://doi.org/}%
\providecommand \selectlanguage [0]{\@gobble}%
\providecommand \bibinfo  [0]{\@secondoftwo}%
\providecommand \bibfield  [0]{\@secondoftwo}%
\providecommand \translation [1]{[#1]}%
\providecommand \BibitemOpen [0]{}%
\providecommand \bibitemStop [0]{}%
\providecommand \bibitemNoStop [0]{.\EOS\space}%
\providecommand \EOS [0]{\spacefactor3000\relax}%
\providecommand \BibitemShut  [1]{\csname bibitem#1\endcsname}%
\let\auto@bib@innerbib\@empty
\bibitem [{\citenamefont {Feynman}()}]{feynman1957the}%
  \BibitemOpen
  \bibfield  {author} {\bibinfo {author} {\bibfnamefont {R.~P.}\ \bibnamefont
  {Feynman}},\ }in\ \href@noop {} {\emph {\bibinfo {booktitle} {The Role of
  Gravitation in Physics. Report from the 1957 Chapel Hill Conference (Max
  Planck Research Library for the History and Development of Knowledge,
  2011)}}},\ \bibinfo {editor} {edited by\ \bibinfo {editor} {\bibfnamefont
  {C.~M.}\ \bibnamefont {DeWitt}}\ and\ \bibinfo {editor} {\bibfnamefont
  {D.}~\bibnamefont {Rickles}}}\BibitemShut {NoStop}%
\bibitem [{\citenamefont {Penrose}(1996)}]{penrose1996gravity}%
  \BibitemOpen
  \bibfield  {author} {\bibinfo {author} {\bibfnamefont {R.}~\bibnamefont
  {Penrose}},\ }\bibfield  {title} {\bibinfo {title} {On gravity's role in
  quantum state reduction},\ }\href
  {https://doi.org/https://doi.org/10.1007/BF02105068} {\bibfield  {journal}
  {\bibinfo  {journal} {General relativity and gravitation}\ }\textbf {\bibinfo
  {volume} {28}},\ \bibinfo {pages} {581} (\bibinfo {year} {1996})}\BibitemShut
  {NoStop}%
\bibitem [{\citenamefont {Di\'osi}(1989)}]{diosi1989models}%
  \BibitemOpen
  \bibfield  {author} {\bibinfo {author} {\bibfnamefont {L.}~\bibnamefont
  {Di\'osi}},\ }\bibfield  {title} {\bibinfo {title} {Models for universal
  reduction of macroscopic quantum fluctuations},\ }\href
  {https://doi.org/10.1103/PhysRevA.40.1165} {\bibfield  {journal} {\bibinfo
  {journal} {Phys. Rev. A}\ }\textbf {\bibinfo {volume} {40}},\ \bibinfo
  {pages} {1165} (\bibinfo {year} {1989})}\BibitemShut {NoStop}%
\bibitem [{\citenamefont {Oppenheim}\ and\ \citenamefont
  {Weller-Davies}(2022)}]{oppenheim2022constraints}%
  \BibitemOpen
  \bibfield  {author} {\bibinfo {author} {\bibfnamefont {J.}~\bibnamefont
  {Oppenheim}}\ and\ \bibinfo {author} {\bibfnamefont {Z.}~\bibnamefont
  {Weller-Davies}},\ }\bibfield  {title} {\bibinfo {title} {The constraints of
  post-quantum classical gravity},\ }\href
  {https://doi.org/https://doi.org/10.1007/JHEP02(2022)080} {\bibfield
  {journal} {\bibinfo  {journal} {Journal of High Energy Physics}\ }\textbf
  {\bibinfo {volume} {2022}},\ \bibinfo {pages} {80} (\bibinfo {year}
  {2022})}\BibitemShut {NoStop}%
\bibitem [{\citenamefont {Dai}\ \emph {et~al.}(2022)\citenamefont {Dai},
  \citenamefont {Minic},\ and\ \citenamefont
  {Stojkovic}}]{10.3389/fphy.2022.891977}%
  \BibitemOpen
  \bibfield  {author} {\bibinfo {author} {\bibfnamefont {D.-C.}\ \bibnamefont
  {Dai}}, \bibinfo {author} {\bibfnamefont {D.}~\bibnamefont {Minic}},\ and\
  \bibinfo {author} {\bibfnamefont {D.}~\bibnamefont {Stojkovic}},\ }\bibfield
  {title} {\bibinfo {title} {On black holes as macroscopic quantum objects},\
  }\bibfield  {journal} {\bibinfo  {journal} {Frontiers in Physics}\ }\textbf
  {\bibinfo {volume} {10}},\ \href {https://doi.org/10.3389/fphy.2022.891977}
  {10.3389/fphy.2022.891977} (\bibinfo {year} {2022})\BibitemShut {NoStop}%
\bibitem [{\citenamefont {Bose}\ \emph {et~al.}(2023)\citenamefont {Bose},
  \citenamefont {Fuentes}, \citenamefont {Geraci}, \citenamefont {Khan},
  \citenamefont {Qvarfort}, \citenamefont {Rademacher}, \citenamefont {Rashid},
  \citenamefont {Toroš}, \citenamefont {Ulbricht},\ and\ \citenamefont
  {Wanjura}}]{bose2023massive}%
  \BibitemOpen
  \bibfield  {author} {\bibinfo {author} {\bibfnamefont {S.}~\bibnamefont
  {Bose}}, \bibinfo {author} {\bibfnamefont {I.}~\bibnamefont {Fuentes}},
  \bibinfo {author} {\bibfnamefont {A.~A.}\ \bibnamefont {Geraci}}, \bibinfo
  {author} {\bibfnamefont {S.~M.}\ \bibnamefont {Khan}}, \bibinfo {author}
  {\bibfnamefont {S.}~\bibnamefont {Qvarfort}}, \bibinfo {author}
  {\bibfnamefont {M.}~\bibnamefont {Rademacher}}, \bibinfo {author}
  {\bibfnamefont {M.}~\bibnamefont {Rashid}}, \bibinfo {author} {\bibfnamefont
  {M.}~\bibnamefont {Toroš}}, \bibinfo {author} {\bibfnamefont
  {H.}~\bibnamefont {Ulbricht}},\ and\ \bibinfo {author} {\bibfnamefont
  {C.~C.}\ \bibnamefont {Wanjura}},\ }\bibfield  {title} {\bibinfo {title}
  {Massive quantum systems as interfaces of quantum mechanics and gravity},\
  }\bibfield  {journal} {\bibinfo  {journal} {arXiv preprint arXiv:2311.09218}\
  }\href {https://doi.org/https://doi.org/10.48550/arXiv.2311.09218}
  {https://doi.org/10.48550/arXiv.2311.09218} (\bibinfo {year}
  {2023})\BibitemShut {NoStop}%
\bibitem [{\citenamefont {Bose}\ \emph {et~al.}(1999)\citenamefont {Bose},
  \citenamefont {Jacobs},\ and\ \citenamefont {Knight}}]{bose1999scheme}%
  \BibitemOpen
  \bibfield  {author} {\bibinfo {author} {\bibfnamefont {S.}~\bibnamefont
  {Bose}}, \bibinfo {author} {\bibfnamefont {K.}~\bibnamefont {Jacobs}},\ and\
  \bibinfo {author} {\bibfnamefont {P.~L.}\ \bibnamefont {Knight}},\ }\bibfield
   {title} {\bibinfo {title} {Scheme to probe the decoherence of a macroscopic
  object},\ }\href {https://doi.org/10.1103/PhysRevA.59.3204} {\bibfield
  {journal} {\bibinfo  {journal} {Phys. Rev. A}\ }\textbf {\bibinfo {volume}
  {59}},\ \bibinfo {pages} {3204} (\bibinfo {year} {1999})}\BibitemShut
  {NoStop}%
\bibitem [{\citenamefont {Scala}\ \emph {et~al.}(2013)\citenamefont {Scala},
  \citenamefont {Kim}, \citenamefont {Morley}, \citenamefont {Barker},\ and\
  \citenamefont {Bose}}]{scala2013matter}%
  \BibitemOpen
  \bibfield  {author} {\bibinfo {author} {\bibfnamefont {M.}~\bibnamefont
  {Scala}}, \bibinfo {author} {\bibfnamefont {M.~S.}\ \bibnamefont {Kim}},
  \bibinfo {author} {\bibfnamefont {G.~W.}\ \bibnamefont {Morley}}, \bibinfo
  {author} {\bibfnamefont {P.~F.}\ \bibnamefont {Barker}},\ and\ \bibinfo
  {author} {\bibfnamefont {S.}~\bibnamefont {Bose}},\ }\bibfield  {title}
  {\bibinfo {title} {Matter-wave interferometry of a levitated thermal
  nano-oscillator induced and probed by a spin},\ }\href
  {https://doi.org/10.1103/PhysRevLett.111.180403} {\bibfield  {journal}
  {\bibinfo  {journal} {Phys. Rev. Lett.}\ }\textbf {\bibinfo {volume} {111}},\
  \bibinfo {pages} {180403} (\bibinfo {year} {2013})}\BibitemShut {NoStop}%
\bibitem [{\citenamefont {Wan}\ \emph {et~al.}(2016)\citenamefont {Wan},
  \citenamefont {Scala}, \citenamefont {Morley}, \citenamefont {Rahman},
  \citenamefont {Ulbricht}, \citenamefont {Bateman}, \citenamefont {Barker},
  \citenamefont {Bose},\ and\ \citenamefont {Kim}}]{wan2016free}%
  \BibitemOpen
  \bibfield  {author} {\bibinfo {author} {\bibfnamefont {C.}~\bibnamefont
  {Wan}}, \bibinfo {author} {\bibfnamefont {M.}~\bibnamefont {Scala}}, \bibinfo
  {author} {\bibfnamefont {G.~W.}\ \bibnamefont {Morley}}, \bibinfo {author}
  {\bibfnamefont {A.~A.}\ \bibnamefont {Rahman}}, \bibinfo {author}
  {\bibfnamefont {H.}~\bibnamefont {Ulbricht}}, \bibinfo {author}
  {\bibfnamefont {J.}~\bibnamefont {Bateman}}, \bibinfo {author} {\bibfnamefont
  {P.~F.}\ \bibnamefont {Barker}}, \bibinfo {author} {\bibfnamefont
  {S.}~\bibnamefont {Bose}},\ and\ \bibinfo {author} {\bibfnamefont {M.~S.}\
  \bibnamefont {Kim}},\ }\bibfield  {title} {\bibinfo {title} {Free nano-object
  ramsey interferometry for large quantum superpositions},\ }\href
  {https://doi.org/10.1103/PhysRevLett.117.143003} {\bibfield  {journal}
  {\bibinfo  {journal} {Phys. Rev. Lett.}\ }\textbf {\bibinfo {volume} {117}},\
  \bibinfo {pages} {143003} (\bibinfo {year} {2016})}\BibitemShut {NoStop}%
\bibitem [{\citenamefont {Pedernales}\ \emph {et~al.}(2020)\citenamefont
  {Pedernales}, \citenamefont {Morley},\ and\ \citenamefont
  {Plenio}}]{pedernales2020motional}%
  \BibitemOpen
  \bibfield  {author} {\bibinfo {author} {\bibfnamefont {J.~S.}\ \bibnamefont
  {Pedernales}}, \bibinfo {author} {\bibfnamefont {G.~W.}\ \bibnamefont
  {Morley}},\ and\ \bibinfo {author} {\bibfnamefont {M.~B.}\ \bibnamefont
  {Plenio}},\ }\bibfield  {title} {\bibinfo {title} {Motional dynamical
  decoupling for interferometry with macroscopic particles},\ }\href
  {https://doi.org/10.1103/PhysRevLett.125.023602} {\bibfield  {journal}
  {\bibinfo  {journal} {Phys. Rev. Lett.}\ }\textbf {\bibinfo {volume} {125}},\
  \bibinfo {pages} {023602} (\bibinfo {year} {2020})}\BibitemShut {NoStop}%
\bibitem [{\citenamefont {Marshman}\ \emph {et~al.}(2022)\citenamefont
  {Marshman}, \citenamefont {Mazumdar}, \citenamefont {Folman},\ and\
  \citenamefont {Bose}}]{marshman2022constructing}%
  \BibitemOpen
  \bibfield  {author} {\bibinfo {author} {\bibfnamefont {R.~J.}\ \bibnamefont
  {Marshman}}, \bibinfo {author} {\bibfnamefont {A.}~\bibnamefont {Mazumdar}},
  \bibinfo {author} {\bibfnamefont {R.}~\bibnamefont {Folman}},\ and\ \bibinfo
  {author} {\bibfnamefont {S.}~\bibnamefont {Bose}},\ }\bibfield  {title}
  {\bibinfo {title} {Constructing nano-object quantum superpositions with a
  stern-gerlach interferometer},\ }\href
  {https://doi.org/10.1103/PhysRevResearch.4.023087} {\bibfield  {journal}
  {\bibinfo  {journal} {Phys. Rev. Res.}\ }\textbf {\bibinfo {volume} {4}},\
  \bibinfo {pages} {023087} (\bibinfo {year} {2022})}\BibitemShut {NoStop}%
\bibitem [{\citenamefont {Margalit}\ \emph {et~al.}(2021)\citenamefont
  {Margalit}, \citenamefont {Dobkowski}, \citenamefont {Zhou}, \citenamefont
  {Amit}, \citenamefont {Japha}, \citenamefont {Moukouri}, \citenamefont
  {Rohrlich}, \citenamefont {Mazumdar}, \citenamefont {Bose}, \citenamefont
  {Henkel},\ and\ \citenamefont {Folman}}]{margalit2020realization}%
  \BibitemOpen
  \bibfield  {author} {\bibinfo {author} {\bibfnamefont {Y.}~\bibnamefont
  {Margalit}}, \bibinfo {author} {\bibfnamefont {O.}~\bibnamefont {Dobkowski}},
  \bibinfo {author} {\bibfnamefont {Z.}~\bibnamefont {Zhou}}, \bibinfo {author}
  {\bibfnamefont {O.}~\bibnamefont {Amit}}, \bibinfo {author} {\bibfnamefont
  {Y.}~\bibnamefont {Japha}}, \bibinfo {author} {\bibfnamefont
  {S.}~\bibnamefont {Moukouri}}, \bibinfo {author} {\bibfnamefont
  {D.}~\bibnamefont {Rohrlich}}, \bibinfo {author} {\bibfnamefont
  {A.}~\bibnamefont {Mazumdar}}, \bibinfo {author} {\bibfnamefont
  {S.}~\bibnamefont {Bose}}, \bibinfo {author} {\bibfnamefont {C.}~\bibnamefont
  {Henkel}},\ and\ \bibinfo {author} {\bibfnamefont {R.}~\bibnamefont
  {Folman}},\ }\bibfield  {title} {\bibinfo {title} {Realization of a complete
  stern-gerlach interferometer: Toward a test of quantum gravity},\ }\href
  {https://doi.org/10.1126/sciadv.abg2879} {\bibfield  {journal} {\bibinfo
  {journal} {Science Advances}\ }\textbf {\bibinfo {volume} {7}},\ \bibinfo
  {pages} {eabg2879} (\bibinfo {year} {2021})},\ \Eprint
  {https://arxiv.org/abs/https://www.science.org/doi/pdf/10.1126/sciadv.abg2879}
  {https://www.science.org/doi/pdf/10.1126/sciadv.abg2879} \BibitemShut
  {NoStop}%
\bibitem [{\citenamefont {Zhou}\ \emph {et~al.}(2022)\citenamefont {Zhou},
  \citenamefont {Marshman}, \citenamefont {Bose},\ and\ \citenamefont
  {Mazumdar}}]{zhou2022catapulting}%
  \BibitemOpen
  \bibfield  {author} {\bibinfo {author} {\bibfnamefont {R.}~\bibnamefont
  {Zhou}}, \bibinfo {author} {\bibfnamefont {R.~J.}\ \bibnamefont {Marshman}},
  \bibinfo {author} {\bibfnamefont {S.}~\bibnamefont {Bose}},\ and\ \bibinfo
  {author} {\bibfnamefont {A.}~\bibnamefont {Mazumdar}},\ }\bibfield  {title}
  {\bibinfo {title} {Catapulting towards massive and large spatial quantum
  superposition},\ }\href {https://doi.org/10.1103/PhysRevResearch.4.043157}
  {\bibfield  {journal} {\bibinfo  {journal} {Phys. Rev. Res.}\ }\textbf
  {\bibinfo {volume} {4}},\ \bibinfo {pages} {043157} (\bibinfo {year}
  {2022})}\BibitemShut {NoStop}%
\bibitem [{\citenamefont {Zhou}\ \emph {et~al.}(2023)\citenamefont {Zhou},
  \citenamefont {Marshman}, \citenamefont {Bose},\ and\ \citenamefont
  {Mazumdar}}]{zhou2022mass}%
  \BibitemOpen
  \bibfield  {author} {\bibinfo {author} {\bibfnamefont {R.}~\bibnamefont
  {Zhou}}, \bibinfo {author} {\bibfnamefont {R.~J.}\ \bibnamefont {Marshman}},
  \bibinfo {author} {\bibfnamefont {S.}~\bibnamefont {Bose}},\ and\ \bibinfo
  {author} {\bibfnamefont {A.}~\bibnamefont {Mazumdar}},\ }\bibfield  {title}
  {\bibinfo {title} {Mass-independent scheme for enhancing spatial quantum
  superpositions},\ }\href {https://doi.org/10.1103/PhysRevA.107.032212}
  {\bibfield  {journal} {\bibinfo  {journal} {Phys. Rev. A}\ }\textbf {\bibinfo
  {volume} {107}},\ \bibinfo {pages} {032212} (\bibinfo {year}
  {2023})}\BibitemShut {NoStop}%
\bibitem [{\citenamefont {Donoghue}(1995)}]{donoghue1995introduction}%
  \BibitemOpen
  \bibfield  {author} {\bibinfo {author} {\bibfnamefont {J.~F.}\ \bibnamefont
  {Donoghue}},\ }\bibfield  {title} {\bibinfo {title} {Introduction to the
  effective field theory description of gravity},\ }\href@noop {} {\bibfield
  {journal} {\bibinfo  {journal} {Advanced school on effective theories:
  Almunecar, Granada, Spain}\ }\textbf {\bibinfo {volume} {26}},\ \bibinfo
  {pages} {217} (\bibinfo {year} {1995})}\BibitemShut {NoStop}%
\bibitem [{\citenamefont {Parikh}\ \emph
  {et~al.}(2021{\natexlab{a}})\citenamefont {Parikh}, \citenamefont {Wilczek},\
  and\ \citenamefont {Zahariade}}]{parikh2021signatures}%
  \BibitemOpen
  \bibfield  {author} {\bibinfo {author} {\bibfnamefont {M.}~\bibnamefont
  {Parikh}}, \bibinfo {author} {\bibfnamefont {F.}~\bibnamefont {Wilczek}},\
  and\ \bibinfo {author} {\bibfnamefont {G.}~\bibnamefont {Zahariade}},\
  }\bibfield  {title} {\bibinfo {title} {Signatures of the quantization of
  gravity at gravitational wave detectors},\ }\href
  {https://doi.org/10.1103/PhysRevD.104.046021} {\bibfield  {journal} {\bibinfo
   {journal} {Phys. Rev. D}\ }\textbf {\bibinfo {volume} {104}},\ \bibinfo
  {pages} {046021} (\bibinfo {year} {2021}{\natexlab{a}})}\BibitemShut
  {NoStop}%
\bibitem [{\citenamefont {Parikh}\ \emph
  {et~al.}(2021{\natexlab{b}})\citenamefont {Parikh}, \citenamefont {Wilczek},\
  and\ \citenamefont {Zahariade}}]{parikh2021quantum}%
  \BibitemOpen
  \bibfield  {author} {\bibinfo {author} {\bibfnamefont {M.}~\bibnamefont
  {Parikh}}, \bibinfo {author} {\bibfnamefont {F.}~\bibnamefont {Wilczek}},\
  and\ \bibinfo {author} {\bibfnamefont {G.}~\bibnamefont {Zahariade}},\
  }\bibfield  {title} {\bibinfo {title} {Quantum mechanics of gravitational
  waves},\ }\href {https://doi.org/10.1103/PhysRevLett.127.081602} {\bibfield
  {journal} {\bibinfo  {journal} {Phys. Rev. Lett.}\ }\textbf {\bibinfo
  {volume} {127}},\ \bibinfo {pages} {081602} (\bibinfo {year}
  {2021}{\natexlab{b}})}\BibitemShut {NoStop}%
\bibitem [{\citenamefont {Tobar}\ \emph {et~al.}(2023)\citenamefont {Tobar},
  \citenamefont {Manikandan}, \citenamefont {Beitel},\ and\ \citenamefont
  {Pikovski}}]{Tobar2023detecting}%
  \BibitemOpen
  \bibfield  {author} {\bibinfo {author} {\bibfnamefont {G.}~\bibnamefont
  {Tobar}}, \bibinfo {author} {\bibfnamefont {S.~K.}\ \bibnamefont
  {Manikandan}}, \bibinfo {author} {\bibfnamefont {T.}~\bibnamefont {Beitel}},\
  and\ \bibinfo {author} {\bibfnamefont {I.}~\bibnamefont {Pikovski}},\
  }\bibfield  {title} {\bibinfo {title} {Detecting single gravitons with
  quantum sensing},\ }\bibfield  {journal} {\bibinfo  {journal} {arXiv preprint
  arXiv:2308.15440}\ }\href
  {https://doi.org/https://doi.org/10.48550/arXiv.2308.15440}
  {https://doi.org/10.48550/arXiv.2308.15440} (\bibinfo {year}
  {2023})\BibitemShut {NoStop}%
\bibitem [{\citenamefont {Carney}\ \emph {et~al.}(2024)\citenamefont {Carney},
  \citenamefont {Domcke},\ and\ \citenamefont {Rodd}}]{Carney2024graviton}%
  \BibitemOpen
  \bibfield  {author} {\bibinfo {author} {\bibfnamefont {D.}~\bibnamefont
  {Carney}}, \bibinfo {author} {\bibfnamefont {V.}~\bibnamefont {Domcke}},\
  and\ \bibinfo {author} {\bibfnamefont {N.~L.}\ \bibnamefont {Rodd}},\
  }\bibfield  {title} {\bibinfo {title} {Graviton detection and the
  quantization of gravity},\ }\href
  {https://doi.org/10.1103/PhysRevD.109.044009} {\bibfield  {journal} {\bibinfo
   {journal} {Phys. Rev. D}\ }\textbf {\bibinfo {volume} {109}},\ \bibinfo
  {pages} {044009} (\bibinfo {year} {2024})}\BibitemShut {NoStop}%
\bibitem [{\citenamefont {Bose}(2016)}]{bose2016matter}%
  \BibitemOpen
  \bibfield  {author} {\bibinfo {author} {\bibfnamefont {S.}~\bibnamefont
  {Bose}},\ }\href@noop {} {\bibinfo {title} {Matter wave ramsey interferometry
  \& the quantum nature of gravity}},\ \bibinfo {howpublished} {Available at
  \url{https://www.youtube.com/watch?v=0Fv-0k13s_k}} (\bibinfo {year} {2016}),\
  \bibinfo {note} {fundamental Problems of Quantum Physics, ICTS,
  Bangalore}\BibitemShut {NoStop}%
\bibitem [{\citenamefont {Bose}\ \emph {et~al.}(2017)\citenamefont {Bose},
  \citenamefont {Mazumdar}, \citenamefont {Morley}, \citenamefont {Ulbricht},
  \citenamefont {Toro\ifmmode~\check{s}\else \v{s}\fi{}}, \citenamefont
  {Paternostro}, \citenamefont {Geraci}, \citenamefont {Barker}, \citenamefont
  {Kim},\ and\ \citenamefont {Milburn}}]{bose2017spin}%
  \BibitemOpen
  \bibfield  {author} {\bibinfo {author} {\bibfnamefont {S.}~\bibnamefont
  {Bose}}, \bibinfo {author} {\bibfnamefont {A.}~\bibnamefont {Mazumdar}},
  \bibinfo {author} {\bibfnamefont {G.~W.}\ \bibnamefont {Morley}}, \bibinfo
  {author} {\bibfnamefont {H.}~\bibnamefont {Ulbricht}}, \bibinfo {author}
  {\bibfnamefont {M.}~\bibnamefont {Toro\ifmmode~\check{s}\else \v{s}\fi{}}},
  \bibinfo {author} {\bibfnamefont {M.}~\bibnamefont {Paternostro}}, \bibinfo
  {author} {\bibfnamefont {A.~A.}\ \bibnamefont {Geraci}}, \bibinfo {author}
  {\bibfnamefont {P.~F.}\ \bibnamefont {Barker}}, \bibinfo {author}
  {\bibfnamefont {M.~S.}\ \bibnamefont {Kim}},\ and\ \bibinfo {author}
  {\bibfnamefont {G.}~\bibnamefont {Milburn}},\ }\bibfield  {title} {\bibinfo
  {title} {Spin entanglement witness for quantum gravity},\ }\href
  {https://doi.org/10.1103/PhysRevLett.119.240401} {\bibfield  {journal}
  {\bibinfo  {journal} {Phys. Rev. Lett.}\ }\textbf {\bibinfo {volume} {119}},\
  \bibinfo {pages} {240401} (\bibinfo {year} {2017})}\BibitemShut {NoStop}%
\bibitem [{\citenamefont {Marletto}\ and\ \citenamefont
  {Vedral}(2017)}]{marletto2017gravitationally}%
  \BibitemOpen
  \bibfield  {author} {\bibinfo {author} {\bibfnamefont {C.}~\bibnamefont
  {Marletto}}\ and\ \bibinfo {author} {\bibfnamefont {V.}~\bibnamefont
  {Vedral}},\ }\bibfield  {title} {\bibinfo {title} {Gravitationally induced
  entanglement between two massive particles is sufficient evidence of quantum
  effects in gravity},\ }\href {https://doi.org/10.1103/PhysRevLett.119.240402}
  {\bibfield  {journal} {\bibinfo  {journal} {Phys. Rev. Lett.}\ }\textbf
  {\bibinfo {volume} {119}},\ \bibinfo {pages} {240402} (\bibinfo {year}
  {2017})}\BibitemShut {NoStop}%
\bibitem [{\citenamefont {Christodoulou}\ and\ \citenamefont
  {Rovelli}(2019)}]{christodoulou2019possibility}%
  \BibitemOpen
  \bibfield  {author} {\bibinfo {author} {\bibfnamefont {M.}~\bibnamefont
  {Christodoulou}}\ and\ \bibinfo {author} {\bibfnamefont {C.}~\bibnamefont
  {Rovelli}},\ }\bibfield  {title} {\bibinfo {title} {On the possibility of
  laboratory evidence for quantum superposition of geometries},\ }\href
  {https://doi.org/https://doi.org/10.1016/j.physletb.2019.03.015} {\bibfield
  {journal} {\bibinfo  {journal} {Physics Letters B}\ }\textbf {\bibinfo
  {volume} {792}},\ \bibinfo {pages} {64} (\bibinfo {year} {2019})}\BibitemShut
  {NoStop}%
\bibitem [{\citenamefont {Marshman}\ \emph {et~al.}(2020)\citenamefont
  {Marshman}, \citenamefont {Mazumdar},\ and\ \citenamefont
  {Bose}}]{marshman2020locality}%
  \BibitemOpen
  \bibfield  {author} {\bibinfo {author} {\bibfnamefont {R.~J.}\ \bibnamefont
  {Marshman}}, \bibinfo {author} {\bibfnamefont {A.}~\bibnamefont {Mazumdar}},\
  and\ \bibinfo {author} {\bibfnamefont {S.}~\bibnamefont {Bose}},\ }\bibfield
  {title} {\bibinfo {title} {Locality and entanglement in table-top testing of
  the quantum nature of linearized gravity},\ }\href
  {https://doi.org/10.1103/PhysRevA.101.052110} {\bibfield  {journal} {\bibinfo
   {journal} {Phys. Rev. A}\ }\textbf {\bibinfo {volume} {101}},\ \bibinfo
  {pages} {052110} (\bibinfo {year} {2020})}\BibitemShut {NoStop}%
\bibitem [{\citenamefont {Bose}\ \emph {et~al.}(2022)\citenamefont {Bose},
  \citenamefont {Mazumdar}, \citenamefont {Schut},\ and\ \citenamefont
  {Toro\ifmmode~\check{s}\else \v{s}\fi{}}}]{bose2022mechanism}%
  \BibitemOpen
  \bibfield  {author} {\bibinfo {author} {\bibfnamefont {S.}~\bibnamefont
  {Bose}}, \bibinfo {author} {\bibfnamefont {A.}~\bibnamefont {Mazumdar}},
  \bibinfo {author} {\bibfnamefont {M.}~\bibnamefont {Schut}},\ and\ \bibinfo
  {author} {\bibfnamefont {M.}~\bibnamefont {Toro\ifmmode~\check{s}\else
  \v{s}\fi{}}},\ }\bibfield  {title} {\bibinfo {title} {Mechanism for the
  quantum natured gravitons to entangle masses},\ }\href
  {https://doi.org/10.1103/PhysRevD.105.106028} {\bibfield  {journal} {\bibinfo
   {journal} {Phys. Rev. D}\ }\textbf {\bibinfo {volume} {105}},\ \bibinfo
  {pages} {106028} (\bibinfo {year} {2022})}\BibitemShut {NoStop}%
\bibitem [{\citenamefont {Belenchia}\ \emph {et~al.}(2018)\citenamefont
  {Belenchia}, \citenamefont {Wald}, \citenamefont {Giacomini}, \citenamefont
  {Castro-Ruiz}, \citenamefont {Brukner},\ and\ \citenamefont
  {Aspelmeyer}}]{belenchia2018quantum}%
  \BibitemOpen
  \bibfield  {author} {\bibinfo {author} {\bibfnamefont {A.}~\bibnamefont
  {Belenchia}}, \bibinfo {author} {\bibfnamefont {R.~M.}\ \bibnamefont {Wald}},
  \bibinfo {author} {\bibfnamefont {F.}~\bibnamefont {Giacomini}}, \bibinfo
  {author} {\bibfnamefont {E.}~\bibnamefont {Castro-Ruiz}}, \bibinfo {author}
  {\bibfnamefont {C.}~\bibnamefont {Brukner}},\ and\ \bibinfo {author}
  {\bibfnamefont {M.}~\bibnamefont {Aspelmeyer}},\ }\bibfield  {title}
  {\bibinfo {title} {Quantum superposition of massive objects and the
  quantization of gravity},\ }\href
  {https://doi.org/10.1103/PhysRevD.98.126009} {\bibfield  {journal} {\bibinfo
  {journal} {Phys. Rev. D}\ }\textbf {\bibinfo {volume} {98}},\ \bibinfo
  {pages} {126009} (\bibinfo {year} {2018})}\BibitemShut {NoStop}%
\bibitem [{\citenamefont {Carney}(2022)}]{carney2022newton}%
  \BibitemOpen
  \bibfield  {author} {\bibinfo {author} {\bibfnamefont {D.}~\bibnamefont
  {Carney}},\ }\bibfield  {title} {\bibinfo {title} {Newton, entanglement, and
  the graviton},\ }\href {https://doi.org/10.1103/PhysRevD.105.024029}
  {\bibfield  {journal} {\bibinfo  {journal} {Phys. Rev. D}\ }\textbf {\bibinfo
  {volume} {105}},\ \bibinfo {pages} {024029} (\bibinfo {year}
  {2022})}\BibitemShut {NoStop}%
\bibitem [{\citenamefont {Danielson}\ \emph {et~al.}(2022)\citenamefont
  {Danielson}, \citenamefont {Satishchandran},\ and\ \citenamefont
  {Wald}}]{danielson2022gravitationally}%
  \BibitemOpen
  \bibfield  {author} {\bibinfo {author} {\bibfnamefont {D.~L.}\ \bibnamefont
  {Danielson}}, \bibinfo {author} {\bibfnamefont {G.}~\bibnamefont
  {Satishchandran}},\ and\ \bibinfo {author} {\bibfnamefont {R.~M.}\
  \bibnamefont {Wald}},\ }\bibfield  {title} {\bibinfo {title} {Gravitationally
  mediated entanglement: Newtonian field versus gravitons},\ }\href
  {https://doi.org/10.1103/PhysRevD.105.086001} {\bibfield  {journal} {\bibinfo
   {journal} {Phys. Rev. D}\ }\textbf {\bibinfo {volume} {105}},\ \bibinfo
  {pages} {086001} (\bibinfo {year} {2022})}\BibitemShut {NoStop}%
\bibitem [{\citenamefont {Galley}\ \emph {et~al.}(2022)\citenamefont {Galley},
  \citenamefont {Giacomini},\ and\ \citenamefont {Selby}}]{galley2022no}%
  \BibitemOpen
  \bibfield  {author} {\bibinfo {author} {\bibfnamefont {T.~D.}\ \bibnamefont
  {Galley}}, \bibinfo {author} {\bibfnamefont {F.}~\bibnamefont {Giacomini}},\
  and\ \bibinfo {author} {\bibfnamefont {J.~H.}\ \bibnamefont {Selby}},\
  }\bibfield  {title} {\bibinfo {title} {A no-go theorem on the nature of the
  gravitational field beyond quantum theory},\ }\href
  {https://doi.org/https://doi.org/10.22331/q-2022-08-17-779} {\bibfield
  {journal} {\bibinfo  {journal} {Quantum}\ }\textbf {\bibinfo {volume} {6}},\
  \bibinfo {pages} {779} (\bibinfo {year} {2022})}\BibitemShut {NoStop}%
\bibitem [{\citenamefont {Christodoulou}\ \emph {et~al.}(2023)\citenamefont
  {Christodoulou}, \citenamefont {Di~Biagio}, \citenamefont {Aspelmeyer},
  \citenamefont {Brukner}, \citenamefont {Rovelli},\ and\ \citenamefont
  {Howl}}]{christodoulou2022locally}%
  \BibitemOpen
  \bibfield  {author} {\bibinfo {author} {\bibfnamefont {M.}~\bibnamefont
  {Christodoulou}}, \bibinfo {author} {\bibfnamefont {A.}~\bibnamefont
  {Di~Biagio}}, \bibinfo {author} {\bibfnamefont {M.}~\bibnamefont
  {Aspelmeyer}}, \bibinfo {author} {\bibfnamefont {C.}~\bibnamefont {Brukner}},
  \bibinfo {author} {\bibfnamefont {C.}~\bibnamefont {Rovelli}},\ and\ \bibinfo
  {author} {\bibfnamefont {R.}~\bibnamefont {Howl}},\ }\bibfield  {title}
  {\bibinfo {title} {Locally mediated entanglement in linearized quantum
  gravity},\ }\href {https://doi.org/10.1103/PhysRevLett.130.100202} {\bibfield
   {journal} {\bibinfo  {journal} {Phys. Rev. Lett.}\ }\textbf {\bibinfo
  {volume} {130}},\ \bibinfo {pages} {100202} (\bibinfo {year}
  {2023})}\BibitemShut {NoStop}%
\bibitem [{\citenamefont {Qvarfort}\ \emph {et~al.}(2020)\citenamefont
  {Qvarfort}, \citenamefont {Bose},\ and\ \citenamefont
  {Serafini}}]{qvarfort2020mesoscopic}%
  \BibitemOpen
  \bibfield  {author} {\bibinfo {author} {\bibfnamefont {S.}~\bibnamefont
  {Qvarfort}}, \bibinfo {author} {\bibfnamefont {S.}~\bibnamefont {Bose}},\
  and\ \bibinfo {author} {\bibfnamefont {A.}~\bibnamefont {Serafini}},\
  }\bibfield  {title} {\bibinfo {title} {Mesoscopic entanglement through
  central--potential interactions},\ }\href
  {https://iopscience.iop.org/article/10.1088/1361-6455/abbe8d} {\bibfield
  {journal} {\bibinfo  {journal} {Journal of Physics B: Atomic, Molecular and
  Optical Physics}\ }\textbf {\bibinfo {volume} {53}},\ \bibinfo {pages}
  {235501} (\bibinfo {year} {2020})}\BibitemShut {NoStop}%
\bibitem [{\citenamefont {Krisnanda}\ \emph {et~al.}(2023)\citenamefont
  {Krisnanda}, \citenamefont {Paterek}, \citenamefont {Paternostro},\ and\
  \citenamefont {Liew}}]{krisnanda2022quantum}%
  \BibitemOpen
  \bibfield  {author} {\bibinfo {author} {\bibfnamefont {T.}~\bibnamefont
  {Krisnanda}}, \bibinfo {author} {\bibfnamefont {T.}~\bibnamefont {Paterek}},
  \bibinfo {author} {\bibfnamefont {M.}~\bibnamefont {Paternostro}},\ and\
  \bibinfo {author} {\bibfnamefont {T.~C.~H.}\ \bibnamefont {Liew}},\
  }\bibfield  {title} {\bibinfo {title} {Quantum neuromorphic approach to
  efficient sensing of gravity-induced entanglement},\ }\href
  {https://doi.org/10.1103/PhysRevD.107.086014} {\bibfield  {journal} {\bibinfo
   {journal} {Phys. Rev. D}\ }\textbf {\bibinfo {volume} {107}},\ \bibinfo
  {pages} {086014} (\bibinfo {year} {2023})}\BibitemShut {NoStop}%
\bibitem [{\citenamefont {Carney}\ \emph {et~al.}(2021)\citenamefont {Carney},
  \citenamefont {M\"uller},\ and\ \citenamefont {Taylor}}]{carney2021using}%
  \BibitemOpen
  \bibfield  {author} {\bibinfo {author} {\bibfnamefont {D.}~\bibnamefont
  {Carney}}, \bibinfo {author} {\bibfnamefont {H.}~\bibnamefont {M\"uller}},\
  and\ \bibinfo {author} {\bibfnamefont {J.~M.}\ \bibnamefont {Taylor}},\
  }\bibfield  {title} {\bibinfo {title} {Using an atom interferometer to infer
  gravitational entanglement generation},\ }\href
  {https://doi.org/10.1103/PRXQuantum.2.030330} {\bibfield  {journal} {\bibinfo
   {journal} {PRX Quantum}\ }\textbf {\bibinfo {volume} {2}},\ \bibinfo {pages}
  {030330} (\bibinfo {year} {2021})}\BibitemShut {NoStop}%
\bibitem [{\citenamefont {Biswas}\ \emph {et~al.}(2023)\citenamefont {Biswas},
  \citenamefont {Bose}, \citenamefont {Mazumdar},\ and\ \citenamefont
  {Toro\ifmmode~\check{s}\else \v{s}\fi{}}}]{Biswas2022Gravitational}%
  \BibitemOpen
  \bibfield  {author} {\bibinfo {author} {\bibfnamefont {D.}~\bibnamefont
  {Biswas}}, \bibinfo {author} {\bibfnamefont {S.}~\bibnamefont {Bose}},
  \bibinfo {author} {\bibfnamefont {A.}~\bibnamefont {Mazumdar}},\ and\
  \bibinfo {author} {\bibfnamefont {M.}~\bibnamefont
  {Toro\ifmmode~\check{s}\else \v{s}\fi{}}},\ }\bibfield  {title} {\bibinfo
  {title} {Gravitational optomechanics: Photon-matter entanglement via graviton
  exchange},\ }\href {https://doi.org/10.1103/PhysRevD.108.064023} {\bibfield
  {journal} {\bibinfo  {journal} {Phys. Rev. D}\ }\textbf {\bibinfo {volume}
  {108}},\ \bibinfo {pages} {064023} (\bibinfo {year} {2023})}\BibitemShut
  {NoStop}%
\bibitem [{\citenamefont {Etezad-Razavi}\ and\ \citenamefont
  {Hardy}(2023)}]{etezad2023paradox}%
  \BibitemOpen
  \bibfield  {author} {\bibinfo {author} {\bibfnamefont {S.}~\bibnamefont
  {Etezad-Razavi}}\ and\ \bibinfo {author} {\bibfnamefont {L.}~\bibnamefont
  {Hardy}},\ }\bibfield  {title} {\bibinfo {title} {Paradox with phase-coupled
  interferometers},\ }\bibfield  {journal} {\bibinfo  {journal} {arXiv preprint
  arXiv:2305.14241}\ }\href
  {https://doi.org/https://doi.org/10.48550/arXiv.2305.14241}
  {https://doi.org/10.48550/arXiv.2305.14241} (\bibinfo {year}
  {2023})\BibitemShut {NoStop}%
\bibitem [{\citenamefont {Kent}\ and\ \citenamefont
  {Pital\'ua-Garc\'{\i}a}(2021)}]{Kent2022bell}%
  \BibitemOpen
  \bibfield  {author} {\bibinfo {author} {\bibfnamefont {A.}~\bibnamefont
  {Kent}}\ and\ \bibinfo {author} {\bibfnamefont {D.}~\bibnamefont
  {Pital\'ua-Garc\'{\i}a}},\ }\bibfield  {title} {\bibinfo {title} {Testing the
  nonclassicality of spacetime: What can we learn from bell--bose et
  al.-marletto-vedral experiments?},\ }\href
  {https://doi.org/10.1103/PhysRevD.104.126030} {\bibfield  {journal} {\bibinfo
   {journal} {Phys. Rev. D}\ }\textbf {\bibinfo {volume} {104}},\ \bibinfo
  {pages} {126030} (\bibinfo {year} {2021})}\BibitemShut {NoStop}%
\bibitem [{\citenamefont {Schut}\ \emph {et~al.}(2022)\citenamefont {Schut},
  \citenamefont {Tilly}, \citenamefont {Marshman}, \citenamefont {Bose},\ and\
  \citenamefont {Mazumdar}}]{schut2022deco}%
  \BibitemOpen
  \bibfield  {author} {\bibinfo {author} {\bibfnamefont {M.}~\bibnamefont
  {Schut}}, \bibinfo {author} {\bibfnamefont {J.}~\bibnamefont {Tilly}},
  \bibinfo {author} {\bibfnamefont {R.~J.}\ \bibnamefont {Marshman}}, \bibinfo
  {author} {\bibfnamefont {S.}~\bibnamefont {Bose}},\ and\ \bibinfo {author}
  {\bibfnamefont {A.}~\bibnamefont {Mazumdar}},\ }\bibfield  {title} {\bibinfo
  {title} {Improving resilience of quantum-gravity-induced entanglement of
  masses to decoherence using three superpositions},\ }\href
  {https://doi.org/10.1103/PhysRevA.105.032411} {\bibfield  {journal} {\bibinfo
   {journal} {Phys. Rev. A}\ }\textbf {\bibinfo {volume} {105}},\ \bibinfo
  {pages} {032411} (\bibinfo {year} {2022})}\BibitemShut {NoStop}%
\bibitem [{\citenamefont {van~de Kamp}\ \emph {et~al.}(2020)\citenamefont
  {van~de Kamp}, \citenamefont {Marshman}, \citenamefont {Bose},\ and\
  \citenamefont {Mazumdar}}]{van2020quantum}%
  \BibitemOpen
  \bibfield  {author} {\bibinfo {author} {\bibfnamefont {T.~W.}\ \bibnamefont
  {van~de Kamp}}, \bibinfo {author} {\bibfnamefont {R.~J.}\ \bibnamefont
  {Marshman}}, \bibinfo {author} {\bibfnamefont {S.}~\bibnamefont {Bose}},\
  and\ \bibinfo {author} {\bibfnamefont {A.}~\bibnamefont {Mazumdar}},\
  }\bibfield  {title} {\bibinfo {title} {Quantum gravity witness via
  entanglement of masses: Casimir screening},\ }\href
  {https://doi.org/10.1103/PhysRevA.102.062807} {\bibfield  {journal} {\bibinfo
   {journal} {Phys. Rev. A}\ }\textbf {\bibinfo {volume} {102}},\ \bibinfo
  {pages} {062807} (\bibinfo {year} {2020})}\BibitemShut {NoStop}%
\bibitem [{\citenamefont {Rijavec}\ \emph {et~al.}(2021)\citenamefont
  {Rijavec}, \citenamefont {Carlesso}, \citenamefont {Bassi}, \citenamefont
  {Vedral},\ and\ \citenamefont {Marletto}}]{Rijavec_2021}%
  \BibitemOpen
  \bibfield  {author} {\bibinfo {author} {\bibfnamefont {S.}~\bibnamefont
  {Rijavec}}, \bibinfo {author} {\bibfnamefont {M.}~\bibnamefont {Carlesso}},
  \bibinfo {author} {\bibfnamefont {A.}~\bibnamefont {Bassi}}, \bibinfo
  {author} {\bibfnamefont {V.}~\bibnamefont {Vedral}},\ and\ \bibinfo {author}
  {\bibfnamefont {C.}~\bibnamefont {Marletto}},\ }\bibfield  {title} {\bibinfo
  {title} {Decoherence effects in non-classicality tests of gravity},\ }\href
  {https://doi.org/10.1088/1367-2630/abf3eb} {\bibfield  {journal} {\bibinfo
  {journal} {New Journal of Physics}\ }\textbf {\bibinfo {volume} {23}},\
  \bibinfo {pages} {043040} (\bibinfo {year} {2021})}\BibitemShut {NoStop}%
\bibitem [{\citenamefont {Peres}(1996)}]{Peres1996sep}%
  \BibitemOpen
  \bibfield  {author} {\bibinfo {author} {\bibfnamefont {A.}~\bibnamefont
  {Peres}},\ }\bibfield  {title} {\bibinfo {title} {Separability criterion for
  density matrices},\ }\href {https://doi.org/10.1103/PhysRevLett.77.1413}
  {\bibfield  {journal} {\bibinfo  {journal} {Phys. Rev. Lett.}\ }\textbf
  {\bibinfo {volume} {77}},\ \bibinfo {pages} {1413} (\bibinfo {year}
  {1996})}\BibitemShut {NoStop}%
\bibitem [{\citenamefont {Horodecki}\ \emph {et~al.}(1996)\citenamefont
  {Horodecki}, \citenamefont {Horodecki},\ and\ \citenamefont
  {Horodecki}}]{HORODECKI19961}%
  \BibitemOpen
  \bibfield  {author} {\bibinfo {author} {\bibfnamefont {M.}~\bibnamefont
  {Horodecki}}, \bibinfo {author} {\bibfnamefont {P.}~\bibnamefont
  {Horodecki}},\ and\ \bibinfo {author} {\bibfnamefont {R.}~\bibnamefont
  {Horodecki}},\ }\bibfield  {title} {\bibinfo {title} {Separability of mixed
  states: necessary and sufficient conditions},\ }\href
  {https://doi.org/https://doi.org/10.1016/S0375-9601(96)00706-2} {\bibfield
  {journal} {\bibinfo  {journal} {Physics Letters A}\ }\textbf {\bibinfo
  {volume} {223}},\ \bibinfo {pages} {1} (\bibinfo {year} {1996})}\BibitemShut
  {NoStop}%
\bibitem [{sup()}]{supp}%
  \BibitemOpen
  \href@noop {} {\bibinfo  {journal} {See the Appendix for the effect of
  decoherence in the earlier entanglement witness based proposal, the examples
  of hybrid models satisfying the NDC, the derivation of the expression of
  quantum violation of the NDC (in the absence as well as in the presence of
  decoherence) in the interferometric setup considered by us, the detailed
  justification for using two probes, estimation of the NDC violations under
  realistic parameter regimes in the absence or in the presence of decoherence.
  This Appendix includes Refs.\cite{HORODECKI1995340,Isart2011quantum}}\
  }\BibitemShut {NoStop}%
\bibitem [{\citenamefont {Nielsen}\ and\ \citenamefont
  {Chuang}(2010)}]{Nielsen_Chuang_2010}%
  \BibitemOpen
\bibfield  {journal} {  }\bibfield  {author} {\bibinfo {author} {\bibfnamefont
  {M.~A.}\ \bibnamefont {Nielsen}}\ and\ \bibinfo {author} {\bibfnamefont
  {I.~L.}\ \bibnamefont {Chuang}},\ }\href@noop {} {\emph {\bibinfo {title}
  {Quantum Computation and Quantum Information: 10th Anniversary Edition}}}\
  (\bibinfo  {publisher} {Cambridge University Press},\ \bibinfo {year}
  {2010})\BibitemShut {NoStop}%
\bibitem [{\citenamefont {Carcassi}\ \emph {et~al.}(2021)\citenamefont
  {Carcassi}, \citenamefont {Maccone},\ and\ \citenamefont
  {Aidala}}]{PhysRevLett.126.110402}%
  \BibitemOpen
  \bibfield  {author} {\bibinfo {author} {\bibfnamefont {G.}~\bibnamefont
  {Carcassi}}, \bibinfo {author} {\bibfnamefont {L.}~\bibnamefont {Maccone}},\
  and\ \bibinfo {author} {\bibfnamefont {C.~A.}\ \bibnamefont {Aidala}},\
  }\bibfield  {title} {\bibinfo {title} {Four postulates of quantum mechanics
  are three},\ }\href {https://doi.org/10.1103/PhysRevLett.126.110402}
  {\bibfield  {journal} {\bibinfo  {journal} {Phys. Rev. Lett.}\ }\textbf
  {\bibinfo {volume} {126}},\ \bibinfo {pages} {110402} (\bibinfo {year}
  {2021})}\BibitemShut {NoStop}%
\bibitem [{\citenamefont {Steane}()}]{steane}%
  \BibitemOpen
  \bibfield  {author} {\bibinfo {author} {\bibfnamefont {A.}~\bibnamefont
  {Steane}},\ }\href@noop {} {}\bibinfo {howpublished}
  {\url{https://users.physics.ox.ac.uk/~Steane/teaching/rel_C_fields.pdf}},\
  \bibinfo {note} {lecture Note (University of Oxford)}\BibitemShut {NoStop}%
\bibitem [{\citenamefont {Leggett}\ and\ \citenamefont
  {Garg}(1985)}]{PhysRevLett.54.857}%
  \BibitemOpen
  \bibfield  {author} {\bibinfo {author} {\bibfnamefont {A.~J.}\ \bibnamefont
  {Leggett}}\ and\ \bibinfo {author} {\bibfnamefont {A.}~\bibnamefont {Garg}},\
  }\bibfield  {title} {\bibinfo {title} {Quantum mechanics versus macroscopic
  realism: Is the flux there when nobody looks?},\ }\href
  {https://doi.org/10.1103/PhysRevLett.54.857} {\bibfield  {journal} {\bibinfo
  {journal} {Phys. Rev. Lett.}\ }\textbf {\bibinfo {volume} {54}},\ \bibinfo
  {pages} {857} (\bibinfo {year} {1985})}\BibitemShut {NoStop}%
\bibitem [{\citenamefont {Knee}\ \emph {et~al.}(2016)\citenamefont {Knee},
  \citenamefont {Kakuyanagi}, \citenamefont {Yeh}, \citenamefont {Matsuzaki},
  \citenamefont {Toida}, \citenamefont {Yamaguchi}, \citenamefont {Saito},
  \citenamefont {Leggett},\ and\ \citenamefont {Munro}}]{NatComm.16}%
  \BibitemOpen
  \bibfield  {author} {\bibinfo {author} {\bibfnamefont {G.~C.}\ \bibnamefont
  {Knee}}, \bibinfo {author} {\bibfnamefont {K.}~\bibnamefont {Kakuyanagi}},
  \bibinfo {author} {\bibfnamefont {M.-C.}\ \bibnamefont {Yeh}}, \bibinfo
  {author} {\bibfnamefont {Y.}~\bibnamefont {Matsuzaki}}, \bibinfo {author}
  {\bibfnamefont {H.}~\bibnamefont {Toida}}, \bibinfo {author} {\bibfnamefont
  {H.}~\bibnamefont {Yamaguchi}}, \bibinfo {author} {\bibfnamefont
  {S.}~\bibnamefont {Saito}}, \bibinfo {author} {\bibfnamefont {A.~J.}\
  \bibnamefont {Leggett}},\ and\ \bibinfo {author} {\bibfnamefont {W.~J.}\
  \bibnamefont {Munro}},\ }\bibfield  {title} {\bibinfo {title} {A strict
  experimental test of macroscopic realism in a superconducting flux qubit},\
  }\href {https://doi.org/https://doi.org/10.1038/ncomms13253} {\bibfield
  {journal} {\bibinfo  {journal} {Nature Communications}\ }\textbf {\bibinfo
  {volume} {7}},\ \bibinfo {pages} {13253} (\bibinfo {year}
  {2016})}\BibitemShut {NoStop}%
\bibitem [{\citenamefont {Kofler}\ and\ \citenamefont
  {Brukner}(2013)}]{PhysRevA.87.052115}%
  \BibitemOpen
  \bibfield  {author} {\bibinfo {author} {\bibfnamefont {J.}~\bibnamefont
  {Kofler}}\ and\ \bibinfo {author} {\bibfnamefont {C.}~\bibnamefont
  {Brukner}},\ }\bibfield  {title} {\bibinfo {title} {Condition for macroscopic
  realism beyond the leggett-garg inequalities},\ }\href
  {https://doi.org/10.1103/PhysRevA.87.052115} {\bibfield  {journal} {\bibinfo
  {journal} {Phys. Rev. A}\ }\textbf {\bibinfo {volume} {87}},\ \bibinfo
  {pages} {052115} (\bibinfo {year} {2013})}\BibitemShut {NoStop}%
\bibitem [{\citenamefont {Schild}\ and\ \citenamefont
  {Emary}(2015)}]{PhysRevA.92.032101}%
  \BibitemOpen
  \bibfield  {author} {\bibinfo {author} {\bibfnamefont {G.}~\bibnamefont
  {Schild}}\ and\ \bibinfo {author} {\bibfnamefont {C.}~\bibnamefont {Emary}},\
  }\bibfield  {title} {\bibinfo {title} {Maximum violations of the
  quantum-witness equality},\ }\href
  {https://doi.org/10.1103/PhysRevA.92.032101} {\bibfield  {journal} {\bibinfo
  {journal} {Phys. Rev. A}\ }\textbf {\bibinfo {volume} {92}},\ \bibinfo
  {pages} {032101} (\bibinfo {year} {2015})}\BibitemShut {NoStop}%
\bibitem [{\citenamefont {Moller}()}]{Moller1962Les}%
  \BibitemOpen
  \bibfield  {author} {\bibinfo {author} {\bibfnamefont {C.}~\bibnamefont
  {Moller}},\ }\bibfield  {title} {\bibinfo {title} {in \textit{Les Theories
  Relativistes de la Gravitation}},\ }\href@noop {} {\bibinfo  {journal}
  {edited by M. A. Lichnerowicz, M. A. Tonnelat (CNRS, Paris, 1962)}\
  }\BibitemShut {NoStop}%
\bibitem [{\citenamefont {Rosenfeld}(1963)}]{ROSENFELD1963353}%
  \BibitemOpen
\bibfield  {journal} {  }\bibfield  {author} {\bibinfo {author} {\bibfnamefont
  {L.}~\bibnamefont {Rosenfeld}},\ }\bibfield  {title} {\bibinfo {title} {On
  quantization of fields},\ }\href
  {https://doi.org/https://doi.org/10.1016/0029-5582(63)90279-7} {\bibfield
  {journal} {\bibinfo  {journal} {Nuclear Physics}\ }\textbf {\bibinfo {volume}
  {40}},\ \bibinfo {pages} {353} (\bibinfo {year} {1963})}\BibitemShut
  {NoStop}%
\bibitem [{\citenamefont {Kafri}\ \emph {et~al.}(2014)\citenamefont {Kafri},
  \citenamefont {Taylor},\ and\ \citenamefont {Milburn}}]{kafri2014classical}%
  \BibitemOpen
  \bibfield  {author} {\bibinfo {author} {\bibfnamefont {D.}~\bibnamefont
  {Kafri}}, \bibinfo {author} {\bibfnamefont {J.}~\bibnamefont {Taylor}},\ and\
  \bibinfo {author} {\bibfnamefont {G.}~\bibnamefont {Milburn}},\ }\bibfield
  {title} {\bibinfo {title} {A classical channel model for gravitational
  decoherence},\ }\href
  {https://iopscience.iop.org/article/10.1088/1367-2630/16/6/065020} {\bibfield
   {journal} {\bibinfo  {journal} {New Journal of Physics}\ }\textbf {\bibinfo
  {volume} {16}},\ \bibinfo {pages} {065020} (\bibinfo {year}
  {2014})}\BibitemShut {NoStop}%
\bibitem [{\citenamefont {Di\'osi}\ and\ \citenamefont
  {Halliwell}(1998)}]{PhysRevLett.81.2846}%
  \BibitemOpen
  \bibfield  {author} {\bibinfo {author} {\bibfnamefont {L.}~\bibnamefont
  {Di\'osi}}\ and\ \bibinfo {author} {\bibfnamefont {J.~J.}\ \bibnamefont
  {Halliwell}},\ }\bibfield  {title} {\bibinfo {title} {Coupling classical and
  quantum variables using continuous quantum measurement theory},\ }\href
  {https://doi.org/10.1103/PhysRevLett.81.2846} {\bibfield  {journal} {\bibinfo
   {journal} {Phys. Rev. Lett.}\ }\textbf {\bibinfo {volume} {81}},\ \bibinfo
  {pages} {2846} (\bibinfo {year} {1998})}\BibitemShut {NoStop}%
\bibitem [{\citenamefont {Page}\ and\ \citenamefont
  {Geilker}(1981)}]{PhysRevLett.47.979}%
  \BibitemOpen
  \bibfield  {author} {\bibinfo {author} {\bibfnamefont {D.~N.}\ \bibnamefont
  {Page}}\ and\ \bibinfo {author} {\bibfnamefont {C.~D.}\ \bibnamefont
  {Geilker}},\ }\bibfield  {title} {\bibinfo {title} {Indirect evidence for
  quantum gravity},\ }\href {https://doi.org/10.1103/PhysRevLett.47.979}
  {\bibfield  {journal} {\bibinfo  {journal} {Phys. Rev. Lett.}\ }\textbf
  {\bibinfo {volume} {47}},\ \bibinfo {pages} {979} (\bibinfo {year}
  {1981})}\BibitemShut {NoStop}%
\bibitem [{\citenamefont {Colella}\ \emph {et~al.}(1975)\citenamefont
  {Colella}, \citenamefont {Overhauser},\ and\ \citenamefont
  {Werner}}]{colella1975observation}%
  \BibitemOpen
  \bibfield  {author} {\bibinfo {author} {\bibfnamefont {R.}~\bibnamefont
  {Colella}}, \bibinfo {author} {\bibfnamefont {A.~W.}\ \bibnamefont
  {Overhauser}},\ and\ \bibinfo {author} {\bibfnamefont {S.~A.}\ \bibnamefont
  {Werner}},\ }\bibfield  {title} {\bibinfo {title} {Observation of
  gravitationally induced quantum interference},\ }\href
  {https://doi.org/10.1103/PhysRevLett.34.1472} {\bibfield  {journal} {\bibinfo
   {journal} {Phys. Rev. Lett.}\ }\textbf {\bibinfo {volume} {34}},\ \bibinfo
  {pages} {1472} (\bibinfo {year} {1975})}\BibitemShut {NoStop}%
\bibitem [{\citenamefont {Schut}\ \emph {et~al.}(2024)\citenamefont {Schut},
  \citenamefont {Geraci}, \citenamefont {Bose},\ and\ \citenamefont
  {Mazumdar}}]{schut2023micron}%
  \BibitemOpen
  \bibfield  {author} {\bibinfo {author} {\bibfnamefont {M.}~\bibnamefont
  {Schut}}, \bibinfo {author} {\bibfnamefont {A.}~\bibnamefont {Geraci}},
  \bibinfo {author} {\bibfnamefont {S.}~\bibnamefont {Bose}},\ and\ \bibinfo
  {author} {\bibfnamefont {A.}~\bibnamefont {Mazumdar}},\ }\bibfield  {title}
  {\bibinfo {title} {Micrometer-size spatial superpositions for the qgem
  protocol via screening and trapping},\ }\href
  {https://doi.org/10.1103/PhysRevResearch.6.013199} {\bibfield  {journal}
  {\bibinfo  {journal} {Phys. Rev. Res.}\ }\textbf {\bibinfo {volume} {6}},\
  \bibinfo {pages} {013199} (\bibinfo {year} {2024})}\BibitemShut {NoStop}%
\bibitem [{\citenamefont {Toro\ifmmode~\check{s}\else \v{s}\fi{}}\ \emph
  {et~al.}(2021)\citenamefont {Toro\ifmmode~\check{s}\else \v{s}\fi{}},
  \citenamefont {van~de Kamp}, \citenamefont {Marshman}, \citenamefont {Kim},
  \citenamefont {Mazumdar},\ and\ \citenamefont {Bose}}]{torovs2020relative}%
  \BibitemOpen
  \bibfield  {author} {\bibinfo {author} {\bibfnamefont {M.}~\bibnamefont
  {Toro\ifmmode~\check{s}\else \v{s}\fi{}}}, \bibinfo {author} {\bibfnamefont
  {T.~W.}\ \bibnamefont {van~de Kamp}}, \bibinfo {author} {\bibfnamefont
  {R.~J.}\ \bibnamefont {Marshman}}, \bibinfo {author} {\bibfnamefont {M.~S.}\
  \bibnamefont {Kim}}, \bibinfo {author} {\bibfnamefont {A.}~\bibnamefont
  {Mazumdar}},\ and\ \bibinfo {author} {\bibfnamefont {S.}~\bibnamefont
  {Bose}},\ }\bibfield  {title} {\bibinfo {title} {Relative acceleration noise
  mitigation for nanocrystal matter-wave interferometry: Applications to
  entangling masses via quantum gravity},\ }\href
  {https://doi.org/10.1103/PhysRevResearch.3.023178} {\bibfield  {journal}
  {\bibinfo  {journal} {Phys. Rev. Res.}\ }\textbf {\bibinfo {volume} {3}},\
  \bibinfo {pages} {023178} (\bibinfo {year} {2021})}\BibitemShut {NoStop}%
\bibitem [{\citenamefont {Bronstein}(2012)}]{Bronstein2012republication}%
  \BibitemOpen
  \bibfield  {author} {\bibinfo {author} {\bibfnamefont {M.}~\bibnamefont
  {Bronstein}},\ }\bibfield  {title} {\bibinfo {title} {Republication of:
  Quantum theory of weak gravitational fields},\ }\href
  {https://doi.org/https://doi.org/10.1007/s10714-011-1285-4} {\bibfield
  {journal} {\bibinfo  {journal} {Gen. Relativ. Gravit.}\ }\textbf {\bibinfo
  {volume} {44}},\ \bibinfo {pages} {267–283} (\bibinfo {year}
  {2012})}\BibitemShut {NoStop}%
\bibitem [{\citenamefont {Torrieri}(2023)}]{Torrieri2023equivalence}%
  \BibitemOpen
  \bibfield  {author} {\bibinfo {author} {\bibfnamefont {G.}~\bibnamefont
  {Torrieri}},\ }\bibfield  {title} {\bibinfo {title} {The equivalence
  principle and inertial-gravitational quantum backreaction},\ }\href
  {https://doi.org/10.1140/epjs/s11734-023-01043-9} {\bibfield  {journal}
  {\bibinfo  {journal} {Eur. Phys. J. Spec. Top.}\ }\textbf {\bibinfo {volume}
  {232}},\ \bibinfo {pages} {3505–3517} (\bibinfo {year} {2023})}\BibitemShut
  {NoStop}%
\bibitem [{tex()}]{text}%
  \BibitemOpen
  \href@noop {} {\bibinfo  {journal} {Any decoherence, on the other hand, can
  be interpreted as a result of the entanglement of the system with the probe
  and/or wider environment; however this entanglement, while being sufficient,
  is not necessary for measurement induced state update. Morover, the
  system-environment entanglement is generically not verifiable because of the
  complexity of the environment.}\ }\BibitemShut {NoStop}%
\bibitem [{\citenamefont {Matsumura}\ \emph {et~al.}(2022)\citenamefont
  {Matsumura}, \citenamefont {Nambu},\ and\ \citenamefont
  {Yamamoto}}]{matsumura2022leggett}%
  \BibitemOpen
\bibfield  {journal} {  }\bibfield  {author} {\bibinfo {author} {\bibfnamefont
  {A.}~\bibnamefont {Matsumura}}, \bibinfo {author} {\bibfnamefont
  {Y.}~\bibnamefont {Nambu}},\ and\ \bibinfo {author} {\bibfnamefont
  {K.}~\bibnamefont {Yamamoto}},\ }\bibfield  {title} {\bibinfo {title}
  {Leggett-garg inequalities for testing quantumness of gravity},\ }\href
  {https://doi.org/10.1103/PhysRevA.106.012214} {\bibfield  {journal} {\bibinfo
   {journal} {Phys. Rev. A}\ }\textbf {\bibinfo {volume} {106}},\ \bibinfo
  {pages} {012214} (\bibinfo {year} {2022})}\BibitemShut {NoStop}%
\bibitem [{\citenamefont {Horodecki}\ \emph {et~al.}(1995)\citenamefont
  {Horodecki}, \citenamefont {Horodecki},\ and\ \citenamefont
  {Horodecki}}]{HORODECKI1995340}%
  \BibitemOpen
  \bibfield  {author} {\bibinfo {author} {\bibfnamefont {R.}~\bibnamefont
  {Horodecki}}, \bibinfo {author} {\bibfnamefont {P.}~\bibnamefont
  {Horodecki}},\ and\ \bibinfo {author} {\bibfnamefont {M.}~\bibnamefont
  {Horodecki}},\ }\bibfield  {title} {\bibinfo {title} {Violating bell
  inequality by mixed spin-12 states: necessary and sufficient condition},\
  }\href {https://doi.org/https://doi.org/10.1016/0375-9601(95)00214-N}
  {\bibfield  {journal} {\bibinfo  {journal} {Physics Letters A}\ }\textbf
  {\bibinfo {volume} {200}},\ \bibinfo {pages} {340} (\bibinfo {year}
  {1995})}\BibitemShut {NoStop}%
\bibitem [{\citenamefont {Romero-Isart}(2011)}]{Isart2011quantum}%
  \BibitemOpen
  \bibfield  {author} {\bibinfo {author} {\bibfnamefont {O.}~\bibnamefont
  {Romero-Isart}},\ }\bibfield  {title} {\bibinfo {title} {Quantum
  superposition of massive objects and collapse models},\ }\href
  {https://doi.org/10.1103/PhysRevA.84.052121} {\bibfield  {journal} {\bibinfo
  {journal} {Phys. Rev. A}\ }\textbf {\bibinfo {volume} {84}},\ \bibinfo
  {pages} {052121} (\bibinfo {year} {2011})}\BibitemShut {NoStop}%
\end{thebibliography}%

\onecolumngrid  


\appendix 

\section{Effect of decoherence in the earlier proposal \cite{bose2016matter,bose2017spin,marletto2017gravitationally}}\label{app00}
\begin{figure}[h]
	\centering
	\includegraphics[width=0.6\textwidth]{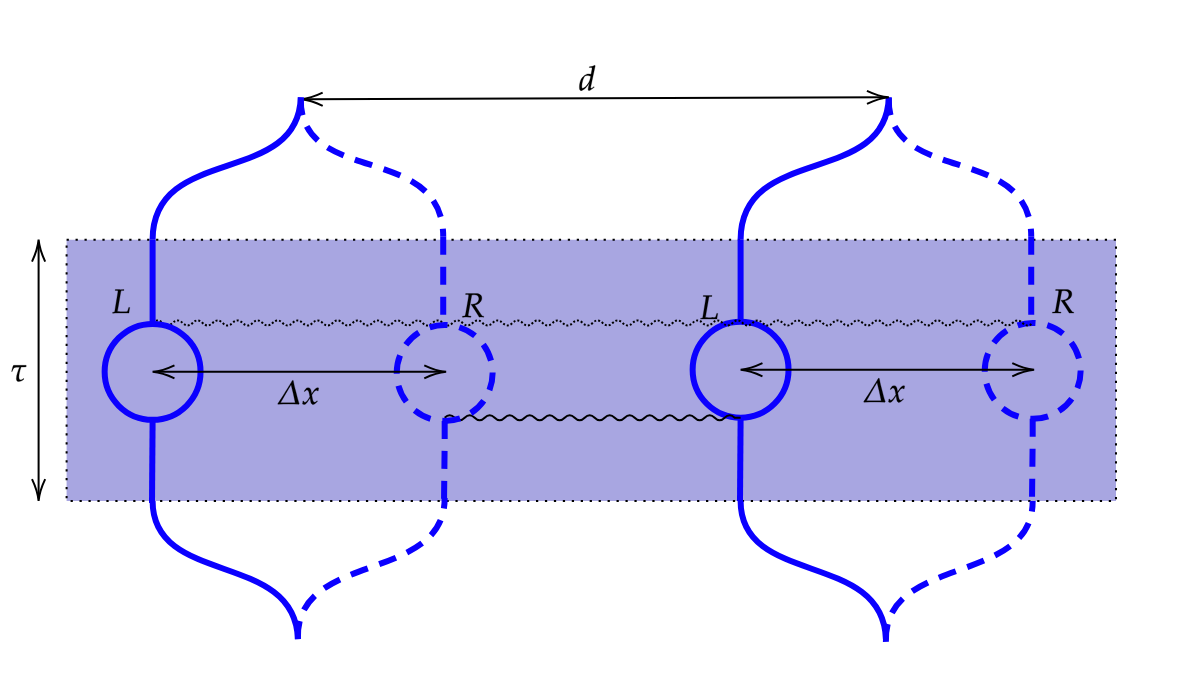}
	\caption{Two masses are prepared in a spatial superposition of width $\Delta x$ through a pair of adjacent interferometers, whose central axes are separated by a distance $d$ \cite{bose2017spin}. The translucent blue region highlights the time $\tau$ for which both superpositions are fully prepared and relative phase development is maximal.}
	\label{fig:QGEM}
\end{figure}

Here, we review the effect of decoherence on the presence of bipartite entanglement between the masses as well as on the detection of entanglement through a violation of the Bell-CHSH (Bell-Clauser-Horne-Shimony-Holt) inequality. We see that entanglement fails to generate altogether for a sufficiently high rate of decoherence $\Gamma$. Moreover, the Bell-CHSH violation has a more stringent requirement on $\Gamma$ than the generation of entanglement itself. This analysis is performed for the original protocol of \cite{bose2016matter,bose2017spin} as depicted in Fig.\ref{fig:QGEM}. For simplicity, we may restrict our attention to the portion of the experiment for which phase development is maximal, i.e., over the duration for which both superpositions have been fully prepared.

Let $A$ and $B$ denote the two massive systems. Let $\ket{\xi_i^{(m)}}$ denotes the environmental state associated with the subsystem state $\ket{i}$ where $i\in\{L,R\}$ and the superscript $m\in\{A,B\}$ denotes the subsystem (system $A$ or system $B$) in question. The embedded spin degrees of freedom have been suppressed for brevity. Upon preparing both superpositions over a time $T$ and neglecting their mutual interactions during that time interval for simplicity, the joint state of the two masses can be taken to be the product state:
\begin{equation}
	\ket{\psi(T)}_{ABE} = \frac{1}{\sqrt{2}}(\ket{L}_A\ket{\xi_{L}^{(A)}}+\ket{R}_A\ket{\xi_{R}^{(A)}}) \otimes \frac{1}{\sqrt{2}}(\ket{L}_B\ket{\xi_{L}^{(B)}}+\ket{R}_B\ket{\xi_{R}^{(B)}}).
\end{equation}
Now considering their gravitational interaction over a time $\tau$, the joint state is given by
\begin{equation}
	\ket{\psi(T+\tau)}_{ABE} = \frac{1}{2}\biggl[\biggl(\ket{L}_A\ket{\xi_{L}^{(A)}} + e^{i\Delta \phi_{RL}}\ket{R}_A\ket{\xi_{R}^{(A)}}\biggr)\ket{L}_B\ket{\xi_{L}^{(B)}} + \biggl(e^{i\Delta \phi_{LR}}\ket{L}_A\ket{\xi_{L}^{(A)}} + \ket{R}_A\ket{\xi_{R}^{(A)}}\biggr)\ket{R}_B\ket{\xi_{R}^{(B)}}\biggr],
\end{equation}
where $\Delta \phi_{RL} = \phi_{RL}-\phi_{LL} = \frac{GMm\tau}{\hbar(d-\Delta x)} - \frac{GMm\tau}{\hbar d}$, and $\Delta \phi_{LR} = \phi_{LR}-\phi_{LL} = \frac{GMm\tau}{\hbar(d+\Delta x)} - \frac{GMm\tau}{\hbar d}$ respectively. Note the overlap of the environmental states $\ket{\xi_i^{(j)}}$ capture the decoherence rate $\Gamma$ through the inner products:
\begin{equation}
	\langle \xi_{i}^{(A)}|\xi_{j}^{(A)}\rangle = \langle \xi_{i}^{(B)}|\xi_{j}^{(B)}\rangle = \eta = e^{- \Gamma \tau}, 
\end{equation}
where $i, j\in \{L,R\}$ with $ i\neq j$. Tracing over the environment, the reduced state of the decohered masses following the interaction is given by,
\begin{equation}\label{densitymatrix}
	\rho_{AB}(T+\tau) = \frac{1}{4}\begin{pmatrix}
		1 & e^{-i\Delta \phi_{LR}}\eta & e^{-i\Delta \phi_{RL}} \eta & \eta^2 \\
		e^{i\Delta \phi_{LR}}\eta & 1 & e^{-i(\Delta \phi_{RL}-\Delta \phi_{LR})}\eta^2 & e^{i\Delta \phi_{LR}} \eta \\
		e^{i\Delta \phi_{RL}} \eta & e^{i(\Delta \phi_{RL}-\Delta \phi_{LR})}\eta^2 & 1 & e^{i\Delta \phi_{RL}} \eta \\
		\eta^2 & e^{-i\Delta \phi_{LR}} \eta & e^{-i\Delta \phi_{RL}} \eta & 1
	\end{pmatrix}.
\end{equation}

\subsection{Entanglement}
The Peres-Horodecki criterion implies that the state of a $2\times 2$ (and $2\times 3$) dimensional system $AB$ is entangled if and only if the partial transpose of the joint density matrix possesses a negative eigenvalue \cite{Peres1996sep,HORODECKI19961}. Computing the partially-transposed state
\begin{equation}
	\rho_{AB}^{T_B}(T+\tau) = \frac{1}{4}\begin{pmatrix}
		1 & e^{i\Delta \phi_{LR}}\eta & e^{-i\Delta \phi_{RL}} \eta & e^{-i(\Delta \phi_{RL}-\Delta \phi_{LR})}\eta^2 \\
		e^{-i\Delta \phi_{LR}}\eta & 1 & \eta^2 & e^{i\Delta \phi_{LR}} \eta \\
		e^{i\Delta \phi_{RL}} \eta & \eta^2 & 1 & e^{-i\Delta \phi_{RL}} \eta \\
		e^{i(\Delta \phi_{RL}-\Delta \phi_{LR})}\eta^2 & e^{-i\Delta \phi_{LR}} \eta & e^{i\Delta \phi_{RL}} \eta & 1
	\end{pmatrix},
\end{equation}
we report the eigenvalues
\begin{align}
	\lambda_1(\tau) & = \frac{1}{2}e^{-\Gamma \tau}\biggl[\sinh \Gamma \tau - \biggl|\sin \overline{\Delta \phi}\biggr|\biggr],\nonumber\\
	\lambda_2(\tau) &= \frac{1}{2}e^{-\Gamma \tau}\biggl[\sinh \Gamma \tau + \biggl|\sin \overline{\Delta \phi}\biggr|\biggr],\nonumber\\
	\lambda_3(\tau) &=\frac{1}{2}e^{-\Gamma \tau}\biggl[\cosh \Gamma \tau - \biggl|\cos \overline{\Delta \phi}\biggr|\biggr],\nonumber\\
	\lambda_4(\tau) &=\frac{1}{2}e^{-\Gamma \tau}\biggl[\cosh \Gamma \tau + \biggl|\cos \overline{\Delta \phi}\biggr|\biggr],
\end{align}
where $\overline{\Delta \phi} =\biggl(\frac{\Delta \phi_{RL} +\Delta \phi_{LR}}{2}\biggr)$ is the averaged relative phase. It is immediate from the ranges of the hyperbolic and trigonometric sines and cosines that $\lambda_2,\lambda_3,\lambda_4 \geq 0$ for all $\tau$, independent of $\Gamma$ and the rates of phase accumulation. Examining $\lambda_1$, it is straightforward to see $\sinh\Gamma \tau \geq \Gamma \tau$, while $|\sin \overline{\Delta \phi}| \leq (d {\overline{\Delta \phi}}/d\tau)\tau$, thus $\lambda_1(\tau) \geq \frac{1}{2}e^{-\Gamma \tau}\biggl[ \Gamma - d {\overline{\Delta \phi}}/d\tau\biggr]\tau$ and all of these bounds are tight. Hence, if $\Gamma$ exceeds the rate of average phase development $d {\overline{\Delta \phi}}/d\tau$, no entanglement between the masses can form.

One can reach similar conclusions in the context of the arrangement of the protocol proposed  in the present article (Fig. \ref{twoprobe}). As before, we neglect the mutual interactions between source and probe masses during generation of superpositions for simplicity, and then consider the joint state of the source and first probe immediately prior to closing of the probe interferometer. The state is identical in form to Eq.\eqref{densitymatrix}, with the redefinitions $\Delta \phi_{LR} = \Delta \phi_{RL} = \frac{GMm\tau}{\hbar\sqrt{d^2+(\Delta x)^2}} - \frac{GMm\tau}{\hbar d} = \Delta \phi$ and $\overline{\Delta \phi} = \Delta \phi$, reflecting the parallel configuration. Thus, we arrive immediately at a similar conclusion for our setup: if $\Gamma$ exceeds $d\Delta \phi/d\tau$, no bipartite entanglement is present immediately prior to measurement, thus demonstrating that the NDC violation persists even in the absence of bipartite entanglement between the source and probe masses.

\subsection{Bell nonlocality}
Horodecki et. al \cite{HORODECKI1995340} gave a necessary and sufficient criterion for a violation of the Bell-CHSH  inequality for any mixed state of two qubits. From the density matrix of Eq.\eqref{densitymatrix}, we first construct the real $3\times 3$ matrix $T_{\rho}$ with entries $T_{mn} = \text{Tr}[\rho_{AB}(T+\tau)(\sigma_m^{A}\otimes \sigma_n^{B})]$, where $\sigma_i$ are the standard single qubit Pauli matrices.
\begin{equation}
{\tiny	T_{\rho} = \begin{pmatrix}
		\eta^2\cos^2\biggl(\frac{\Delta\phi_{RL}-\Delta\phi_{LR}}{2}\biggr) & -\eta^2\cos\biggl(\frac{\Delta\phi_{RL}-\Delta\phi_{LR}}{2}\biggr)\sin\biggl(\frac{\Delta\phi_{RL}-\Delta\phi_{LR}}{2}\biggr) & - \eta \sin\biggl(\frac{\Delta\phi_{RL}+\Delta\phi_{LR}}{2}\biggr)\sin\biggl(\frac{\Delta\phi_{RL}-\Delta\phi_{LR}}{2}\biggr)\\
		\eta^2\cos\biggl(\frac{\Delta\phi_{RL}-\Delta\phi_{LR}}{2}\biggr)\sin\biggl(\frac{\Delta\phi_{RL}-\Delta\phi_{LR}}{2}\biggr)&-\eta^2\sin^2\biggl(\frac{\Delta\phi_{RL}-\Delta\phi_{LR}}{2}\biggr) & \eta \sin\biggl(\frac{\Delta\phi_{RL}+\Delta\phi_{LR}}{2}\biggr)\cos\biggl(\frac{\Delta\phi_{RL}-\Delta\phi_{LR}}{2}\biggr)\\
		\eta \sin\biggl(\frac{\Delta\phi_{RL}+\Delta\phi_{LR}}{2}\biggr)\sin\biggl(\frac{\Delta\phi_{RL}-\Delta\phi_{LR}}{2}\biggr) &\eta \sin\biggl(\frac{\Delta\phi_{RL}+\Delta\phi_{LR}}{2}\biggr)\cos\biggl(\frac{\Delta\phi_{RL}-\Delta\phi_{LR}}{2}\biggr) & 0
	\end{pmatrix}.}
\end{equation} 
Next, we compute the symmetric matrix $U_{\rho} = T_{\rho}^{T}T_{\rho}$.
From this, we note that the eigenvalues of $U_{\rho}$ are:
\begin{align}
	u_1(\tau) &= \eta^4\\
	u_2(\tau)& =u_3(\tau) = \eta^2\sin^2\biggl(\frac{\Delta\phi_{RL}+\Delta\phi_{LR}}{2}\biggr).
\end{align}
It is shown in \cite{HORODECKI1995340} that the Bell-CHSH inequality is violated by the state with density matrix $\rho$ if and only if the sum $M(\rho)$ of the largest pair of the eigenvalues of $U_{\rho}$ exceeds $1$. We consider two cases:
\begin{itemize}
	\item If $\eta \geq |\sin \overline{\Delta \phi}|$, then $M(\rho) = u_1 + u_2 = \eta^2(\eta^2 + \sin^2 \overline{\Delta \phi})$. Thus $M(\rho)>1 \iff \sin^2(\overline{\Delta \phi}) > 2\sinh 2\Gamma \tau$. Seeking a tangential linear upper bound through the origin, one finds $\sin^2(\overline{\Delta \phi}) \lesssim 0.7246 \frac{d\overline{\Delta \phi}}{d\tau}\tau$. Similarly, it is easy to see that $2\sinh 2\Gamma \tau \geq 4\Gamma \tau$. Thus, for a Bell violation, it is necessary for rates of phase development and decoherence to satisfy $0.7246\frac{d\overline{\Delta \phi}}{d\tau} \gtrsim 4\Gamma$. Hence, if $\Gamma \gtrsim 0.1812\frac{d\overline{\Delta \phi}}{d\tau}$, the state fails to violate the Bell CHSH inequality.
	\item If $\eta < |\sin \overline{\Delta \phi}|$, then $M(\rho) = u_2 + u_3 = 2e^{-2\Gamma \tau} \sin ^2\overline{\Delta \phi}$. Viewed as function of $\tau$, $M(\rho)$ is maximised at the first local maximum, occurring at $\tau_*=\biggl(\frac{d\overline{\Delta \phi}}{d\tau}\biggr)^{-1}\tan^{-1}\biggl(\frac{d\overline{\Delta \phi}/d\tau}{\Gamma}\biggr)$. Writing $r=\frac{d\overline{\Delta \phi}/d\tau}{\Gamma}$ as the ratio of rate of phase development to the rate of decoherence, we thus find the following necessary condition for $M(\rho)>1$:
	\begin{equation}
		\frac{2r^2\exp \biggl(-\frac{2}{r}\tan^{-1}r\biggr)}{1+r^2}>1.
	\end{equation}
	Numerically, this implies a violation of the Bell CHSH inequality is attainable only if $r\gtrsim 4.1913$. Thus, if $\Gamma \gtrsim 0.2386 \frac{d\overline{\Delta \phi}}{d\tau}$, the state fails to violate the Bell CHSH inequality. 
\end{itemize}
Notably, these requirements on $\Gamma$ are more demanding than those for the mere presence of bipartite entanglement, illustrating the more stringent constraints on environmental conditions necessary to obtain a Bell violation.
\section{NDC satisfaction examples in hybrid models}\label{app01}

Here ``unmeasured case'' will imply the case where gravitational field is not measured before the final detection of the source
mass in the $+$ and $-$ outputs of the interferometer, and ``measured case'' will imply the case where gravitational field is measured before the final measurement. Now, we present two extreme instances of hybrid models satisfying NDC: 

(i) A Moller-Rosenfeld \cite{Moller1962Les,ROSENFELD1963353} mean field model in which the Newtonian gravitational potential is produced by the expectation value of the mass distribution $\langle T_{00} \rangle$.  In such model, for the measured case,  the probe will simply read out a gravitational potential defined by the average mass distribution 
and thus does not cause any disturbance. Thus $P_{+}$ remains same in both the measured and unmeasured cases, and there is no violation of the NDC condition. 

(ii) Any hybrid model in which gravitational field has a definite state with some probability \cite{oppenheim2022constraints,kafri2014classical,PhysRevLett.81.2846}.  For example, one such model involves spontaneous collapse of the matter wave function \cite{penrose1996gravity}, implying the associated gravitational field acquiring different definite values with different probabilities even in the absence of any measurement. The intermediate measurement just reveals these definite values. 
Thus, again, the $P_{+}$ probability becomes equal in both the measured and unmeasured cases and no violation of the NDC condition is obtained.

\section{Deriving the expression of quantum violation of the NDC}\label{app1}
Let the state-subscripts $M_i$ and $S_i$ denote the mass and the embedded spin degrees of freedom of a given system labelled by $i$ according to whether it is one of the two probes ($i=A,B$) or the source  ($i=C$). We further assume that $M$ is the mass of the source and $m$ is the mass of each of the probes. We first detail the case where the intermediate measurements of the gravitational field by the two probes are done, and then contrast this case with the case where only the final source-spin measurement is performed. The discussion presented in this section is based on Fig. \ref{twoprobe}.

The initial state of the source mass with embedded spin (system 1) at $t=0$ is given by,
\begin{equation}
	\ket{\psi(0)}_{C} = \ket{\zeta}_{M_C}\otimes\dfrac{1}{\sqrt 2}(\ket{\uparrow}_{S_C}+\ket{\downarrow}_{S_C}),
\end{equation}
where $\ket{\zeta}_{M_C}$ is the initial localised state of the source mass at the center of
the axis of the source-interferometer.

Over a time $T$, the source mass  is prepared in spatial superposition via the unitary evolution:
\begin{align}
	\ket{\zeta}_{M_C}\otimes\ket{\uparrow}_{S_C} &\to \ket{L \uparrow}_{C}, \nonumber \\
	\ket{\zeta}_{M_C}\otimes\ket{\downarrow}_{S_C} &\to \ket{R \uparrow}_{C},
	\label{evo}
\end{align}
where the centers of $\ket{L \uparrow}_{C}$ and $\ket{R \uparrow}_{C}$ are separated by a distance $\Delta x$ after time $T$.

Upon completion of this stage, a probe mass  with embedded spin (system $A$) is introduced. The joint state of system $C$ and system $A$ at $t=T$ is given by,
\begin{equation}
	\ket{\psi(T)}_{C,A} = \dfrac{1}{\sqrt 2}(\ket{L\uparrow}_{C}+\ket{R\downarrow}_{C})\otimes \ket{C}_{M_A} \otimes \dfrac{1}{\sqrt 2}(\ket{\uparrow}_{S_A}+\ket{\downarrow}_{S_A}).
\end{equation}

The probe mass now enters into a spatial superposition equivalently to the evolution of Eq.(\ref{evo}) (with the subscript `$C$' in Eq.(\ref{evo}) being replaced by `$A$'), except it also continually acquires gravitational phases due to the interaction with the source mass. Let us suppose that the superposition size (in our case, the distance between the two arms of the interferometer) of the probe system a time $t\in[0,T]$ later is given by the real function $\Delta x(t)$. At any given time $t$, let $d_{xy}(t)$ denote the time-dependent distance between arm $x$ and arm $y$ of the source and probe interferometers respectively, where $x,y\in\{ L,R \}$. These may be written in the following form for the configuration considered in this setup,
\begin{align}
	d_{LL}(t) &= d_{RR}(t) = \sqrt{d^2+\biggl(\dfrac{\Delta x-\Delta x(t)}{2}\biggr)^2},\\
	d_{LR}(t) &= d_{RL}(t) = \sqrt{d^2+\biggl(\dfrac{\Delta x +\Delta x(t)}{2}\biggr)^2}.
\end{align}

At time $T$ later, the probe spatial superposition is now of size $\Delta x(t=T) = \Delta x$, equal to that of the source spatial superposition. Once the probe spatial superposition is fully prepared, the joint state of system $C$ and system $A$ at instant $t=2T$ is given by (discarding an overall phase),
\begin{equation}    \ket{\psi(2T)}_{C,A}=\dfrac{1}{2}\biggl(\ket{L\uparrow}_{C}\ket{L\uparrow}_A + e^{i\Delta \phi_T}\ket{L\uparrow}_{C}\ket{R\downarrow}_A + e^{i\Delta \phi_T}\ket{R\downarrow}_{C}\ket{L\uparrow}_A + \ket{R\downarrow}_{C}\ket{R\downarrow}_A\biggr),
\end{equation}
where 
\begin{equation}
	\Delta \phi_T = \dfrac{GMm}{\hbar d}\int_0^{T} \left(\dfrac{1}{\sqrt{1+\biggl(\dfrac{\Delta x + \Delta x(t)}{2d}\biggr)^2}} - \dfrac{1}{\sqrt{1+\biggl(\dfrac{\Delta x - \Delta x(t)}{2d}\biggr)^2}}\right)dt.\label{supophase}
\end{equation}

The source and probe now interact through gravity in a static arrangement for a time interval $\tau$ before the spatial probe superposition is closed over a time $T$ through the reversal of Eq.\eqref{evo} with the subscript `$C$' being replaced by `$A$'. During closing of the spatial superposition also, $\Delta x(t)$ denotes the superposition size at any instant $t$. Subsequently,  a projective measurement of the probe spin is performed in the $\ket{\pm}_{S_A}=(\ket{\uparrow}_{S_A}\pm \ket{\downarrow}_{S_A})/\sqrt{2}$ basis. As explained in the main paper, this measurement is nothing but a measurement of the source's gravitational field. The joint state of the source and probe prior to this measurement is given by,
\begin{equation}
	\ket{\psi(3T+\tau)}_{C,A}=\dfrac{1}{2}\biggl((\ket{L\uparrow}_{C}+e^{i\Delta \phi}\ket{R\downarrow}_{C})\ket{\uparrow}_{S_A} + (e^{i\Delta \phi}\ket{L\uparrow}_{C}  + \ket{R\downarrow}_{C})\ket{\downarrow}_{S_A}\biggr)\otimes \ket{\zeta}_{M_A},
	\label{3tplustau}
\end{equation}
where an overall phase has again been discarded, and
\begin{equation}
	\Delta \phi = \Delta \phi_{\tau} + 2\Delta \phi_T, \quad \Delta \phi_{\tau} =\dfrac{GMm\tau}{\hbar\sqrt{d^2 + (\Delta x)^2}} - \dfrac{GMm\tau}{\hbar d}=\dfrac{GMm\tau}{\hbar d}\biggl(\dfrac{1}{\sqrt{1 + (\Delta x/d)^2}}-1\biggr).\label{phase}
\end{equation}
Note that $\Delta \phi < 0$. Equivalently, the above state (\ref{3tplustau}) can be expressed as
\begin{align}
	\ket{\psi(3T+\tau)}_{C,A} =&\dfrac{1}{\sqrt{2}}\biggl(\sqrt{1+ \cos \Delta \phi} \dfrac{\left(1 + e^{i\Delta \phi} \right) \ket{L\uparrow}_{C}+\left(e^{i\Delta \phi} + 1\right)\ket{R\downarrow}_{C}}{2\sqrt{1 + \cos \Delta \phi}}\ket{+}_{S_A} \nonumber \\
	& \hspace{0.3cm} + \sqrt{1- \cos \Delta \phi} \dfrac{\left(1 - e^{i\Delta \phi} \right) \ket{L\uparrow}_{C}+\left(e^{i\Delta \phi} -  1\right)\ket{R\downarrow}_{C}}{2\sqrt{1-\cos \Delta \phi}} \ket{-}_{S_A}\biggr) \ket{\zeta}_{M_A}.
	\label{source-probe-ent-app}
\end{align}

Performing the aforementioned projective measurement in the $\ket{\pm}_{S_A}$ basis on the probe and obtaining the outcomes $\pm$, the unnormalised post-measurement states are given by,
\begin{equation}
	\ket{\psi_{\pm}(3T+\tau)}_{C,A}=\dfrac{1}{2\sqrt{2}}\biggl((1\pm e^{i\Delta \phi})\ket{L\uparrow}_{C} + (e^{i\Delta \phi}\pm 1)\ket{R\downarrow}_C\biggr)\ket{\zeta \pm}_A,
\end{equation}

The probe (system $A$) is now decoupled and discarded, and a new probe (system $B$) is immediately introduced. Hence, the joint state of system $C$ and system $B$ at $t= 3 T + \tau$ is given by,
\begin{equation}
	\ket{\psi_{\pm}(3T+\tau)}_{C,B}=\dfrac{1}{2\sqrt{2}}\biggl((1\pm e^{i\Delta \phi})\ket{L\uparrow}_{C} + (e^{i\Delta \phi}\pm 1)\ket{R\downarrow}_C\biggr)\otimes \ket{\zeta}_{M_B}\otimes\dfrac{1}{\sqrt{2}}\biggl(\ket{\uparrow}_{S_B} + \ket{\downarrow}_{S_B}\biggr).
\end{equation}

As before, the new probe now interacts with the source system via the gravitational field in a similar way for a further time $2T+\tau$ (over the course of preparing the probe in a spatial superposition over a time interval of $T$ with $\Delta x (t)$ being the size of the spatial superposition at any time $t$, keeping the fully prepared spatial superposition with size $\Delta x$ in a static configuration for a time $\tau$, and closing the spatial superposition for a further time $T$ with $\Delta x (t)$ again being the spatial superposition size at any instant $t$) before a projective measurement in the $\ket{\pm}_{S_B} = (\ket{\uparrow}_{S_B}\pm \ket{\downarrow}_{S_B})/\sqrt{2}$ basis is performed on the internal spin degree of freedom of system $B$. Effectively, this is also a measurement of the gravity of the source. Ignoring overall phases, the joint state of system $C$ and system $B$ prior to this measurement is given by,
\begin{align}
	\ket{\psi_{\pm}( t_1)}_{C,B}=&\dfrac{1}{4}\biggl[\biggl((1\pm e^{i\Delta \phi})\ket{L\uparrow}_{C}+(e^{i\Delta \phi}\pm 1)e^{i\Delta \phi}\ket{R\downarrow}_{C}\biggr)\ket{\uparrow}_{S_B}\nonumber\\
	&+\biggl((1\pm e^{i\Delta \phi})e^{i\Delta \phi}\ket{L\uparrow}_{C} + (e^{i\Delta \phi}\pm 1)\ket{R\downarrow}_C\biggr)\ket{\downarrow}_{S_B}\biggr]\otimes \ket{\zeta}_{M_B},
\end{align}
where $\Delta \phi$ is as per Eq.\eqref{phase} and $ t_1=5T+2\tau$ is the total time that has elapsed once the two probe measurements have been completed. Consequently, the measurement on the second probe in the $\ket{\pm}_{S_B}$ basis results in the following unnormalised states conditioned on the outcomes of the two measurements,
\begin{equation}
	\ket{\psi_{\pm_1, \pm_2}( t_1)}_{C,B} = \dfrac{1}{4\sqrt{2}}\biggl[\biggl(1\pm_1 e^{i\Delta \phi}\biggr)\biggl(1\pm_2 e^{i\Delta \phi}\biggr)\ket{L\uparrow}_{C} + \biggl(e^{i\Delta \phi}\pm_1 1\biggr)\biggl(e^{i\Delta \phi}\pm_2 1\biggr)\ket{R\downarrow}_{C}\biggr]\ket{\zeta \pm_2}_B,\label{post1}
\end{equation}
where the subscript $\pm_{1}$ denotes the outcome of the first probe (system $A$) measurement, and $\pm_{2}$ denotes the outcome of the second probe (system $B$) measurement.

The second probe is now also discarded, and over a time $t_2-t_1=T$, the spatial superposition of the source interferometer is closed via the reversal of the unitary evolution in Eq.\eqref{evo}. A final projective measurement of the embedded spin state of the source is performed in the $\ket{\pm}_{S_C} = (\ket{\uparrow}_{S_C}\pm \ket{\downarrow}_{S_C})/\sqrt{2}$ basis, yielding the final unnormalised states conditioned on the outcomes of the three measurements,
\begin{equation}    
	\ket{\psi_{\pm_1,\pm_2,\pm_3}(t_2)}_{C} = \dfrac{1}{8}\biggr[\biggl(1\pm_1 e^{i\Delta \phi}\biggr)\biggl(1\pm_2 e^{i\Delta \phi}\biggr) \pm_3 \biggl(e^{i\Delta \phi}\pm_1 1\biggr)\biggl(e^{i\Delta \phi}\pm_2 1\biggr) \biggr]\ket{\zeta \pm_3}_{C},
\end{equation}
where $\pm_{3}$ denotes the outcome of the third, i.e., the final measurement on the spin of the source (system $C$).



Computing the norms of these states, the joint probabilities of these outcomes are given by,
\begin{align}
	P_{+,+,+} =  \langle\psi_{+,+,+}|\psi_{+,+,+}\rangle &= \cos^4\biggl(\dfrac{\Delta \phi}{2}\biggr)\label{P+++1},\\
	P_{-,-,+} = \langle\psi_{-,-,+}|\psi_{-,-,+}\rangle &= \sin^4\biggl(\dfrac{\Delta \phi}{2}\biggr)\label{P--+1},\\
	P_{+,-,-} = \langle\psi_{+,-,-}|\psi_{+,-,-}\rangle &= \sin^2\biggl(\dfrac{\Delta \phi}{2}\biggr)\cos^2 \biggl(\dfrac{\Delta \phi}{2}\biggr)\label{P+--1},\\
	P_{-,+,-} = \langle\psi_{-,+,-}|\psi_{-,+,-}\rangle &= \sin^2\biggl(\dfrac{\Delta \phi}{2}\biggr)\cos^2 \biggl(\dfrac{\Delta \phi}{2}\biggr)\label{P-+-1},\\
	P_{+,-,+}=P_{-,+,+}=P_{+,+,-} &=P_{-,-,-} = 0,\label{Prest1}
\end{align}
where $P_{a,b,c}$ denote the joint probability of getting the outcomes $a$ for the first measurement (measurement of the source's gravitational field by the first probe), $b$ for the second measurement (measurement of the source's gravitational field by the second probe), and $c$ for the third measurement (spin measurement of the source).

Now let us consider the case where the probes are not introduced, and therefore no measurements are performed prior to the final measurement on the system 1 at $t=t_2$. The initial state of the system $C$ at $t=0$ is given by,
\begin{equation}
	\ket{\psi(0)}_{C}  = \ket{\zeta}_{M_C}\otimes\dfrac{1}{\sqrt 2}(\ket{\uparrow}_{S_C}+\ket{\downarrow}_{S_C}).
\end{equation}
A spatial superposition is generated over a time $T$ according to the evolution Eq.\eqref{evo}. For total parity with the first case (involving the aforementioned measurements by the two probes), the system is then held in superposition until  $t=t_1= 5T+2\tau$. The spatial superposition is then unitarily closed over a further time $T$ according to the reversal of Eq.\eqref{evo}. Since there is no probe mass in the vicinity, no gravitational phases are developed throughout this process and the state at time $t=t_2$ is thus simply the initial state of the system,
\begin{equation}
	\ket{\psi(t_2)}_{C} = \ket{\zeta}_{M_C}\otimes\dfrac{1}{\sqrt 2}(\ket{\uparrow}_{S_C}+\ket{\downarrow}_{S_C}).
\end{equation}

At this time, a measurement of the embedded spin is performed in $\ket{\pm}_{S_C}=(\ket{\uparrow}_{S_C}+\ket{\downarrow}_{S_C})/\sqrt{2}$ basis. The unnormalised final post-measurement states conditioned on the outcomes $\pm$ are given by,
\begin{equation}
	\ket{\psi_{\pm}(t_2)}_{C} = \dfrac{1}{2}(1\pm 1)\ket{\zeta \pm}_{C}.
\end{equation}

Tacitly, the probabilities  are straightforwardly seen to be
\begin{align}
	P_{+}=1,\\
	P_{-}=0.
\end{align}

Thus the violation of the NDC, as quantified by the difference between the statistics in the intermediately measured and intermediately unmeasured cases detailed above, is given by,
\begin{equation}
	V(\pm) = P_{\pm} - \displaystyle\sum_{a,b\in\{+,-\}}P_{a,b,\pm} = \pm\dfrac{1}{2}\sin^2\Delta \phi.
	\label{Vpm1}
\end{equation}

\section{Justification for using two probes}\label{app2}

In order to close the loophole of observing a violation of the NDC in the absence of measurement induced disturbance, we should only permit the following stochastic rotation (a consequence of decoherence solely due to the intermediate quantum measurements) on the state of the source mass -- $[(1 + e^{- \beta t})/2 \, \, \mathbb{I} + (1 - e^{- \beta t})/2 \, \, \sigma_z]$, instead of the following -- $\mathcal{R}_z (\theta) [(1 + e^{- \beta t})/2 \, \, \mathbb{I} + (1 - e^{- \beta t})/2 \, \, \sigma_z]$ (here $t$ denotes  the total time-scale of gravitational field measurement by the two probes and $\beta$ denote the rate of quantum measurement-induced decoherence of the source), which is a stochastic rotation with an additional deterministic rotation about $z$-axis by some angle $\theta$. This deterministic rotation due to the presence of the probe is a classical disturbance, and it is independent of whether any quantum measurement process has occurred. Below we show that the above criteria is satisfied in the two-probe setup, as opposed to what is seen in the case of a single probe. We remark that there may be other techniques to eliminate the effect of such classical disturbances, while here we have used this double probe setup as one simple feasible solution.

The reduced density matrix of the source mass with embedded spin after gravitational interactions with the two probes (averaging over all outcomes of the two sequential measurements by the two probes) is given by,
\begin{align}
	\rho_C = \frac{1}{2}\left(1+\cos^2 \Delta \phi \right) \frac{|L \uparrow \rangle_C + |R \downarrow\rangle_C}{\sqrt{2}} \frac{\langle L \uparrow |_C + \langle R \downarrow|_C}{\sqrt{2}} + \frac{1}{2} \left(1-\cos^2 \Delta \phi \right) \frac{|L \uparrow \rangle_C - |R \downarrow\rangle_C}{\sqrt{2}} \frac{\langle L \uparrow |_C - \langle R \downarrow|_C}{\sqrt{2}}.
\end{align}
This is equivalent to the action of the following stochastic rotation on the source mass: $[(1 + e^{- \beta t})/2 \, \, \mathbb{I} + (1 - e^{- \beta t})/2 \, \, \sigma_z]$, where $\beta t = - \text{log}_e \left(\cos^2 \Delta \phi\right)$. This is possible under quantum measurements of the gravitational field (entanglement formation between the source and the each of the two probes, followed by the projection operations on each of the two probes). 

On the other hand, consider the case where a single probe is measured after a total run time $3T+\tau$ equivalently to the above analysis, but instead that the second probe is never introduced and the source superposition is closed over a time $T$ before it's own measurement is performed. In this case, the reduced density matrix of the source mass with embedded spin after gravitational interaction with a single probe (averaging over the outcomes of the probe-measurement) is given by,
\begin{align}
	\rho_C = \frac{1}{2}\left(1+\cos \Delta \phi \right) \frac{|L \uparrow \rangle_C + |R \downarrow\rangle_C}{\sqrt{2}} \frac{\langle L \uparrow |_C + \langle R \downarrow|_C}{\sqrt{2}} + \frac{1}{2} \left(1-\cos \Delta \phi \right) \frac{|L \uparrow \rangle_C - |R \downarrow\rangle_C}{\sqrt{2}} \frac{\langle L \uparrow |_C - \langle R \downarrow|_C}{\sqrt{2}},
	\label{single}
\end{align}
For $0 < \Delta \phi \leq \pi/2$ or for $3\pi/2 \leq \Delta \phi \leq 2\pi$, the above is equivalent to the action of the following stochastic rotation on the source mass: $[(1 + e^{- \beta_1 t})/2 \, \, \mathbb{I} + (1 - e^{- \beta_1 t})/2 \, \, \sigma_z]$, where $\beta_1 t = - \text{log}_e \left(\cos \Delta \phi\right)$. However, for $\pi/2 < \Delta \phi < 3\pi/2$, (\ref{single}) is equivalent to the action of the following combination of deterministic as well as stochastic rotations on the source mass: $\mathcal{R}_z (\theta) [(1 + e^{- \beta_2 t})/2 \, \, \mathbb{I} + (1 - e^{- \beta_2 t})/2 \, \, \sigma_z]$ with $\mathcal{R}_z (\theta) = \sigma_z$, where $\beta_2 t = - \text{log}_e \left(-\cos \Delta \phi\right)$. Hence, in this regime of $\Delta \phi$, the disturbance on the state of the source mass is not solely due to the intermediate quantum measurements. Consequently, this extra disturbance can give rise to a false violation of the NDC, which cannot be interpreted as a consequence of quantum measurement-induced collapse solely. This is the reason we have used two probes instead of a single probe.

In light of the above discussion, the following can be interpreted. In case of two probes as considered by us, the action on the source mass due to the first probe is given by, $\mathcal{R}_z (\theta) [(1 + e^{- \beta_i t})/2 \, \, \mathbb{I} + (1 - e^{- \beta_i t})/2 \, \, \sigma_z]$ and the action on the source mass due to the second probe is given by, $\mathcal{R}_z (\theta) [(1 + e^{- \beta_i t})/2 \, \, \mathbb{I} + (1 - e^{- \beta_i t})/2 \, \, \sigma_z]$, where $\mathcal{R}_z (\theta) = \mathbb{I}$ when $0 < \Delta \phi \leq \pi/2$ or $3\pi/2 \leq \Delta \phi \leq 2\pi$ and $\mathcal{R}_z (\theta) = \sigma_z$, when $\pi/2 < \Delta \phi < 3\pi/2$; $i=1$  when $0 < \Delta \phi \leq \pi/2$ or $3\pi/2 \leq \Delta \phi \leq 2\pi$ and $i=2$ when $\pi/2 < \Delta \phi < 3\pi/2$; $\beta_1 t = - \text{log}_e \left(\cos \Delta \phi\right)$ and $\beta_2 t = - \text{log}_e \left(-\cos \Delta \phi\right)$. Therefore, the total action on the source mass is given by, $\mathcal{R}_z (\theta) [(1 + e^{- \beta_i t})/2 \, \, \mathbb{I} + (1 - e^{- \beta_i t})/2 \, \, \sigma_z] \, \, \mathcal{R}_z (\theta) [(1 + e^{- \beta_i t})/2 \, \, \mathbb{I} + (1 - e^{- \beta_i t})/2 \, \, \sigma_z]$, which, for any $\Delta \phi$, equals to 
\begin{align}
	&\mathcal{R}_z (\theta) \left[\frac{1 + e^{- \beta_i t}}{2} \, \, \mathbb{I} + \frac{1 - e^{- \beta_i t}}{2} \, \, \sigma_z \right] \, \, \mathcal{R}_z (\theta) \left[\frac{1 + e^{- \beta_i t}}{2} \, \, \mathbb{I} + \frac{1 - e^{- \beta_i t}}{2} \, \, \sigma_z\right] \nonumber \\
	&=\left[\frac{1 + e^{- \beta_i t}}{2} \, \, \mathbb{I} + \frac{1 - e^{- \beta_i t}}{2} \, \, \sigma_z\right] \, \, \mathcal{R}_z (\theta) \, \, \mathcal{R}_z (\theta) \left[\frac{1 + e^{- \beta_i t}}{2} \, \, \mathbb{I} + \frac{1 - e^{- \beta_i t}}{2} \, \, \sigma_z\right] \nonumber \\
	&= \left[\frac{1 + e^{- \beta_i t}}{2} \, \, \mathbb{I} + \frac{1 - e^{- \beta_i t}}{2} \, \, \sigma_z\right]  \left[\frac{1 + e^{- \beta_i t}}{2} \, \, \mathbb{I} + \frac{1 - e^{- \beta_i t}}{2} \, \, \sigma_z \right] \nonumber \\
	&= \frac{1 + e^{- 2\beta_i t}}{2}\, \, \mathbb{I} + \frac{1 - e^{- 2\beta_i t}}{2} \, \, \sigma_z \nonumber \\
	&= \frac{1 +\cos^2 \Delta \phi}{2}\, \, \mathbb{I} + \frac{1 - \cos^2 \Delta \phi}{2} \, \, \sigma_z.
\end{align}
Hence, effect of the deterministic rotation due  to the first probe is eliminated by the effect of the deterministic rotation due  to the second probe as $\mathcal{R}_z (\theta)\mathcal{R}_z (\theta) = \mathbb{I}$ for $\mathcal{R}_z (\theta) = \mathbb{I}$ as well as for $\mathcal{R}_z (\theta) = \sigma_z$.

It may be noted that the single probe setup does not give rise to any deterministic rotation for $0 < \Delta \phi \leq \pi/2$ or for $3\pi/2 \leq \Delta \phi \leq 2\pi$. Hence, in this range of $\Delta \phi$, violation of the NDC with a single probe can also be used as a signature of quantum measurement-induced disturbance. That is, our proposal can be implemented with a single probe to test quantum nature of gravity as long as $0 < \Delta \phi \leq \pi/2$ or $3\pi/2 \leq \Delta \phi \leq 2\pi$. However, such a constraint on $\Delta \phi$ restricts the choice of the relevant parameters in experimental context, which can be difficult to achieve in reality. On the other hand, the double-probe setup considered by us does not impose any such constraint on $\Delta \phi$, which gives a lot of freedom in choosing the relevant experimental parameters.

\section{Deriving the expression of quantum violation of the NDC in the presence of decoherence}\label{app4}
With all quantum processes involving the preparation and maintenance of quantum states, it is crucial to account for the ever-present effect of decoherence due to interaction with environment. In this section, we will treat this issue with care. Here, we account for this by coupling all system states $\ket{i}$ to environmental states $\ket{\xi_i}$ with overlaps between different  $\ket{\xi_i}$ decaying exponentially according to a rate of decoherence $\Gamma$ fixed by the external environmental conditions over the duration of the presence of the various systems \cite{Isart2011quantum}. Note, typically, that $\Gamma$ depends exclusively on the details of the evolution of the superposition size $\Delta x(t)$. Putting in  another way, given the dynamics of the superposition growth, the decoherence function $\Gamma$ will now be a non-negative functional of the superposition size, $\Gamma [\Delta x(t)]$. From this perspective, the overlap of environmental states associated with decoherence during the preparation of a superposition (of size $\Delta x$, achieved over a time $T$, say) will be given by,

\begin{equation}
	\langle \xi_i|\xi_j\rangle = \exp \biggl(-\int_0^T \Gamma [\Delta x(t)]dt\biggr) \, \, \forall i \neq j.
\end{equation}

Assuming reversal symmetry in the trajectory of superposition generation while closing a superposition, the overlaps are similarly given by,
\begin{equation}
	\langle \xi_i|\xi_j\rangle = \exp \biggl(-\int_0^T \Gamma [\Delta x(T-t)]dt\biggr) \, \, \forall i \neq j.
\end{equation}

We hence observe that the decoherence due to opening or closing the superposition is  the same regardless of which is being done, provided sufficient reversal symmetry. 
Let $\ket{\xi_i^{(m)}}$ denote the environmental state associated with the subsystem state $\ket{i}$ where $i\in\{L,R\}$ and the superscript $m\in\{A,B,C\}$ denotes the subsystem in question, i.e., source (system $C$), first probe (system $A$) or second probe (system $B$) respectively. Given that we have already understood the evolution of the states of all subsystems involved in the previous section, one can quite easily foreshadow the overlap of environmental states that will arise by the end of the full experiment to quantify their decoherence effects as follows
\begin{align}
	\eta_1 &=\langle \xi_i^{(C)}|\xi_j^{(C)} \rangle =\exp \biggl(-2\int_0^{T} \Gamma [\Delta x(t)]dt -\Gamma_{\text{Max}}(4T+2\tau)\biggr) \, \, \forall i \neq j,\label{deco3}\\ 
	\eta_2 &= \langle \xi_i^{(A)}|\xi_j^{(A)} \rangle =\langle \xi_i^{(B)}|\xi_j^{(B)} \rangle = \exp \biggl(-2\int_0^T \Gamma [\Delta x(t)]dt - \Gamma_{\text{Max}} \tau\biggr) \, \, \forall i \neq j,\label{deco2}
\end{align}
where $\Gamma_{\text{Max}}=\Gamma[\Delta x(t=T)]$ is the maximum rate of decoherence, applicable when the superposition size has reached $\Delta x(t=T) = \Delta x$. 

Recounting the details: At first, the source mass  (with embedded spin where the spin degrees of freedom being initially in a superposition) is prepared in a spatial superposition in a time interval $T$. Then the first probe (mass with embedded spin, where the spin degrees of freedom is in superposition) is introduced and its mass degrees of freedom is prepared in a spatial superposition in another time interval $T$. The probe mass and the source mass then interact in static formation for a time $\tau$, which is followed by closing of the spatial superposition of the probe in a time interval of $T$. Finally, measurement on the probe in the $\ket{\pm}_{S_A} = (\ket{\uparrow}_{S_A}\pm \ket{\downarrow}_{S_A})/\sqrt{2}$ basis is performed. The unnormalised post-measurement state of the source system and the first probe is given by,
\begin{align}
	\ket{\psi_{\pm}(3T+\tau)}_{C,A} = \dfrac{1}{2\sqrt{2}}\biggl[&\biggl(\ket{\xi_L^{(A)}}\pm e^{i\Delta \phi}\ket{\xi_R^{(A)}}\biggr)\ket{L\uparrow}_{C}\ket{\xi_L^{(C)}} + \biggl(e^{i\Delta \phi}\ket{\xi_L^{(A)}}\pm \ket{\xi_R^{(A)}}\biggr)\ket{R\downarrow}_{C}\ket{\xi_R^{(C)}}\biggr] \otimes \ket{\zeta \pm}_{A},
\end{align}
where $\Delta \phi$ is defined as per Eq.\eqref{phase}. 

The first probe is then discarded and the second probe is immediately introduced. As before, the second probe now interacts with the source system via the gravitational field for a time $2T+\tau$ (over the course of preparing the spatial probe superposition over a time $T$, keeping the fully prepared spatial superposition in static arrangement for a time $\tau$ and closing the spatial superposition over a time $T$) before a projective measurement in the $\ket{\pm}_{S_B} = (\ket{\uparrow}_{S_B}\pm \ket{\downarrow}_{S_B})/\sqrt{2}$ basis is performed. Ignoring overall phases, the unnormalised joint state of the second probe and source system immediately prior to the projective spin measurement is given by (where $t_1 = 5T+2\tau$),
\begin{align}
	\ket{\psi_{\pm}(t_1)}_{C,B}=\dfrac{1}{4}&\biggl[\biggl(\ket{\xi_L^{(A)}}\pm e^{i\Delta \phi}\ket{\xi_R^{(A)}}\biggr)\ket{L\uparrow}_{C}\ket{\uparrow}_{S_B}\ket{\xi_L^{(C)}}\ket{\xi_L^{(B)}}+\biggl(\ket{\xi_L^{(A)}}\pm e^{i\Delta \phi}\ket{\xi_R^{(A)}}\biggr)e^{i\Delta \phi}\ket{L\uparrow}_{C}\ket{\downarrow}_{S_B}\ket{\xi_L^{(C)}}\ket{\xi_R^{(B)}}\nonumber\\
	&+\biggl(e^{i\Delta \phi}\ket{\xi_L^{(A)}}\pm \ket{\xi_R^{(A)}}\biggr)e^{i\Delta \phi}\ket{R\downarrow}_{C}\ket{\uparrow}_{S_B}\ket{\xi_R^{(C)}}\ket{\xi_L^{(B)}} + \biggl(e^{i\Delta \phi}\ket{\xi_L^{(A)}}\pm \ket{\xi_R^{(A)}}\biggr)\ket{R\downarrow}_{C}\ket{\downarrow}_{S_B}\ket{\xi_R^{(C)}}\ket{\xi_R^{(B)}} \biggr] \otimes \ket{\zeta}_{M_B}.
\end{align}

Completing the measurement of the gravitational field under the second probe results in the following unnormalised states conditioned on the outcomes of the two measurements,

\begin{align}
	\ket{\psi_{\pm_1, \pm_2}(t_1)}_{C,B} = \dfrac{1}{4\sqrt{2}}\biggl[&\biggl(\ket{\xi_L^{(A)}}\pm_1 e^{i\Delta \phi}\ket{\xi_R^{(A)}}\biggr)\biggl(\ket{\xi_L^{(B)}}\pm_2 e^{i\Delta \phi}\ket{\xi_R^{(B)}}\biggr)\ket{L\uparrow}_{C}\ket{\xi_L^{(C)}} \nonumber\\
	&+ \biggl(e^{i\Delta \phi}\ket{\xi_L^{(A)}}\pm_1 \ket{\xi_R^{(A)}}\biggr)\biggl(e^{i\Delta \phi}\ket{\xi_L^{(B)}}\pm_2 \ket{\xi_R^{(B)}}\biggr)\ket{R\downarrow}_C\ket{\xi_R^{(C)}}\biggr]\ket{\zeta \pm_2}_B.\label{post2}
\end{align}

At this stage, the second probe  is discarded, and during a time interval $t_2 - t_1=T$ the spatial superposition of the source mass is closed, and then a measurement of the embedded spin of the source (system $C$) in $\ket{\pm}_{S_C}=(\ket{\uparrow}_{S_C}+\ket{\downarrow}_{S_C})/\sqrt{2}$ basis is performed. The final unnormalised states conditioned on all possible measurement outcomes are therefore given by,
\begin{align}
	\ket{\psi_{\pm_1,\pm_2,\pm_3}(t_2)}_{1} = \dfrac{1}{8}\biggl[&\biggl(\ket{\xi_L^{(A)}}\pm_1 e^{i\Delta \phi}\ket{\xi_R^{(A)}}\biggr)\biggl(\ket{\xi_L^{(B)}}\pm_2 e^{i\Delta \phi}\ket{\xi_R^{(B)}}\biggr)\ket{\xi_L^{(C)}} \nonumber\\
	&\pm_3 \biggl(e^{i\Delta \phi}\ket{\xi_L^{(A)}}\pm_1 \ket{\xi_R^{(A)}}\biggr)\biggl(e^{i\Delta \phi}\ket{\xi_L^{(B)}}\pm_2 \ket{\xi_R^{(B)}}\biggr)\ket{\xi_R^{(C)}}\biggr]\ket{\zeta \pm_3}_{C}.
\end{align}

Computing the norms of these eight states thus yields the joint probabilities for all combinations of measurement outcomes,
\begin{align}
	P_{+++} &= \dfrac{1}{8}\biggl[\biggl(1+\eta_2 \cos\Delta\phi\biggr)^2 + \eta_1 \biggl(\cos\Delta\phi + \eta_2\biggr)^2\biggr],\label{P+++2}\\
	P_{+-+} &= \dfrac{1}{8}\biggl[1-\eta_2 ^2 \cos^2\Delta \phi + \eta_1 \biggl(\cos^2\Delta\phi - \eta_2 ^2 \biggr)\biggr],\label{P+-+2}\\
	P_{-++} &= \dfrac{1}{8}\biggl[1-\eta_2^2\cos^2\Delta \phi + \eta_1 \biggl(\cos^2\Delta\phi - \eta_2 ^2 \biggr)\biggr],\label{P-++2}\\
	P_{--+} & =\dfrac{1}{8}\biggl[\biggl(1-\eta_2 \cos\Delta\phi\biggr)^2 + \eta_1\biggl(\cos\Delta\phi - \eta_2)^2\biggr],\label{P--+2}\\
	P_{++-} & =\dfrac{1}{8}\biggl[\biggl(1+\eta_2 \cos\Delta\phi\biggr)^2 - \eta_1\biggl(\cos\Delta\phi +  \eta_2 \biggr)^2\biggr],\label{P++-2}\\
	P_{+--} & = \dfrac{1}{8}\biggl[1-\eta_2 ^2 \cos^2\Delta \phi - \eta_1 \biggl(\cos^2\Delta\phi - \eta_2^2 \biggr)\biggr],\label{P+--2}\\
	P_{-+-} & = \dfrac{1}{8}\biggl[1-\eta_2^2\cos^2\Delta \phi - \eta_1 \biggl(\cos^2\Delta\phi - \eta_2 ^2\biggr)\biggr],\label{P-+-2}\\
	P_{---} & =\dfrac{1}{8}\biggl[\biggl(1-\eta_2 \cos\Delta\phi\biggr)^2 - \eta_1\biggl(\cos\Delta\phi - \eta_2 \biggr)^2\biggr],\label{P---2}
\end{align}
where $\eta_1,\eta_2$ are defined according to the overlaps of Eqs.\eqref{deco3} and \eqref{deco2}. 

Next, considering the effect of decoherence in the case of no intermediate measurements of the gravitational field, note the initial state of the system is given by,
\begin{equation}
	\ket{\psi(0)}_{C} = \ket{\zeta}_{M_C}\otimes \dfrac{1}{\sqrt{2}}\biggl(\ket{\uparrow}_{S_C}+\ket{\downarrow}_{S_C}\biggr).
\end{equation}

Performing the same protocol without the introduction of probes over a time interval $t_2$ and performing a projective measurement of the internal spin degree of freedom in $\ket{\pm}_{S_C}=(\ket{\uparrow}_{S_C}+\ket{\downarrow}_{S_C})/\sqrt{2}$ basis with outcomes $\pm$ yields the following final unnormalized states,
\begin{equation}
	\ket{\psi_{\pm}(t_2)}_{C} = \dfrac{1}{2}\biggl(\ket{\xi_L^{(C)}}\pm\ket{\xi_R^{(C)}}\biggr)\ket{\zeta \pm}_C.
\end{equation}

Taking the norm of these states yields the probabilities of outcomes $\pm$ given that no intermediate measurements have been performed:
\begin{equation}
	P_{\pm} = \dfrac{1}{2}\biggl(1\pm \eta_1\biggr).
\end{equation}

Note, summing the appropriate probabilities, we get
\begin{equation}
	\displaystyle\sum_{a,b\in\{+,-\}}P_{ab\pm} = \dfrac{1}{2}\biggl(1\pm \eta_1 \cos^2\Delta \phi\biggr).
\end{equation}
Hence, the violation of the NDC scales with the overlaps of the environmental states as follows
\begin{equation}
	V(\pm) = P_{\pm} - \displaystyle\sum_{a,b\in\{+,-\}}P_{ab\pm} = 
	\pm\dfrac{1}{2}\eta_1\sin^2\Delta \phi =  \pm\dfrac{1}{2}\exp \biggl(-2\int_0^T \Gamma [\Delta x(t)]dt -\Gamma_{\text{Max}} (4T+2\tau)\biggr)\sin^2\Delta \phi.
\end{equation}

By construction, $\Delta x(t)$ is taken to be maximised at $t=T$ as it first achieves the desired final superposition size $\Delta x = \Delta x(t=T)$ at that time. Assuming reasonably that $\Gamma$ is monotonically increasing as a function of a monotonically increasing $\Delta x(t)$, it therefore achieves its maximum at $t=T$ as well. Hence, using the maximal bound, the integral in the exponent above is crudely bounded above by $\Gamma_{\text{Max}} T$, where $\Gamma_{\text{Max}} = \Gamma[\Delta x(t=T)]$. Thus
\begin{equation}
	\dfrac{1}{2}e^{-\Gamma_{\text{Max}}t_2}\sin^2{\Delta \phi} \leq  |V(\pm)| \leq \dfrac{1}{2}\sin^2{\Delta \phi},
	\label{decobound}
\end{equation}
where the upper bound corresponds to the decoherence-free case. On the other hand, the lower bound physically reflects the case of maximal overestimation of decoherence effects during the preparation and closing of the spatial superpositions by assuming that $\Gamma[\Delta x(t)] = \Gamma_{\text{Max}} $ for all $t$, whereas in reality $\Gamma[\Delta x(t)] \leq \Gamma_{\text{Max}}$ for all $t$.

\begin{figure}[t!]
	\centering
	\includegraphics[width=14cm,height=14cm]{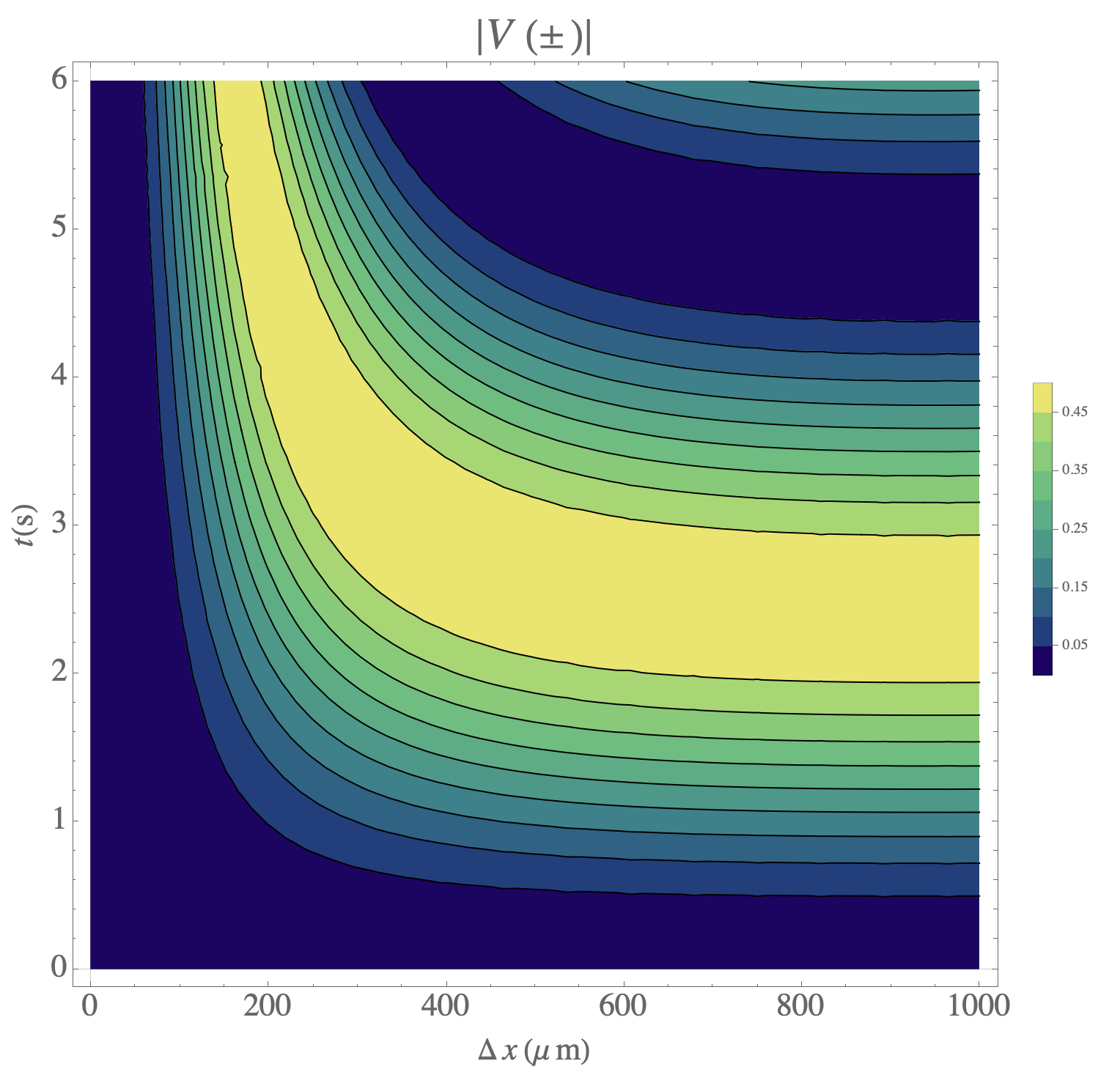}
	\caption{Variation of NDC violations according to Eq.\eqref{Vpm1} with $T= \tau = t$ versus $\Delta x$ and $t$ for $d\sim 157\mu$m in the limit of negligible environmental decoherence. As the time $t$ increases to approximately $2.2$s, a growing and eventually maximal violation of the NDC emerges for a wide range of superposition widths $\Delta x$. For $t\geq 3$s (approx.), the range of superposition sizes $\Delta x$ for which the maximum NDC violation of $0.5$ can be obtained becomes narrower.}\label{fig:violation2}
\end{figure}

Notably, when the decoherence rate exceeds a critical value defined by the rate of relative phase accumulation, the gravity induced entanglement witness protocol \cite{bose2016matter,bose2017spin,marletto2017gravitationally} is no longer effective \cite{schut2022deco,van2020quantum,Rijavec_2021}. However, a quantum violation of the NDC persists in the present protocol for any decoherence rate. 
This is a consequence of the fact that the joint state of the source-probes-environment remains entangled for any decoherence rate (implying the possibility of disturbance of the gravitational field due to measurements by the probes), whereas the reduced state of the source-probes (after tracing out the environment) becomes separable.

\section{Estimation of violations under realistic parameter regimes}\label{app5}


\begin{figure}[t!]
	\centering
	\includegraphics[width=14cm,height=14cm]{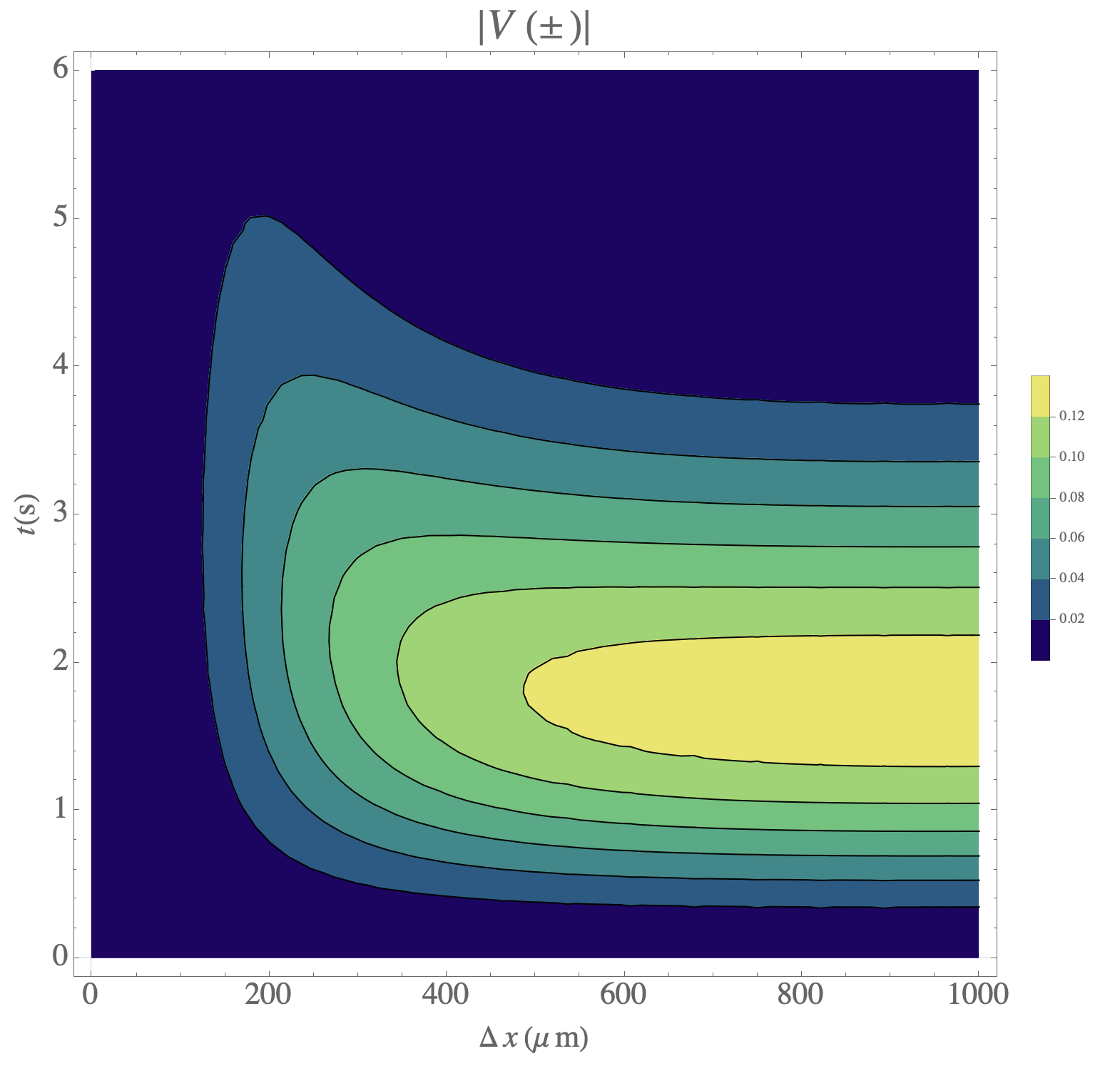}
	\caption{Behaviour of the lower bound for NDC violations (defined by Eq.\eqref{decobound} with $t_2 = 6 T + 2 \tau = 8 t$, where we have taken $T= \tau = t$) for $d\sim 157\mu$m in the presence of decoherence at a maximum rate $\Gamma_{\text{Max}}\sim 0.08$Hz. All violations are  damped by an overestimated factor of $\exp(-8\Gamma_{\text{Max}}t)$, but violations obtained through higher interaction times are punished more strongly by decoherence effects.}
	\label{fig:Fig-3-DecoViolations}
\end{figure}

In this section, we consider the expression for the total phase accumulation given by Eq.\eqref{phase} in terms of physical parameters, where we choose a simple model for the formation/closing of the spatial superposition of the probes that is linear-in-time:
\begin{equation}
	\Delta x(t) = \frac{\Delta x}{T}t.
\end{equation}
Under this assumption, the integral expression \eqref{supophase} for the phase developed during the opening and closing of probe superpositions may be evaluated analytically, and we find that
\begin{equation}
	\Delta \phi_T = \dfrac{GMmT}{\hbar d} \dfrac{2d}{\Delta x}\log\left[\dfrac{\frac{\Delta x}{d} + \sqrt{1+(\frac{\Delta x}{d})^2}}{\biggl(\frac{\Delta x}{2d} + \sqrt{1+(\frac{\Delta x}{2d})^2}\biggr)^2}\right].
\end{equation}
Thus, combining this with Eq.\eqref{phase}, the total relative phase accumulated  is given by,
\begin{equation}
	\Delta \phi = \dfrac{GMm}{\hbar d}\left(\left(\dfrac{1}{\sqrt{ 1+(\frac{\Delta x}{d})^2}}-1\right)\tau + \dfrac{4d}{\Delta x}\log\left[\dfrac{\frac{\Delta x}{d} + \sqrt{1+(\frac{\Delta x}{d})^2}}{\biggl(\frac{\Delta x}{2d} + \sqrt{1+(\frac{\Delta x}{2d})^2}\biggr)^2}\right]T\right)\label{phase2}.
\end{equation}
For simplicity, let us suppose all masses are taken to be identical, $M=m\sim 10^{-14}$kg, and that timescale $\tau$ is of similar order to $T$: $\tau = T = t$. In this case, Eq.\eqref{phase2} and thus the violation given by Eq.\eqref{Vpm1} are fully specified by the value of two dimensionless quantities--
\begin{equation}
	\alpha = \dfrac{Gm^2t}{\hbar d}, \qquad \beta = \dfrac{\Delta x}{d}.
\end{equation}

In practice, the choice of the distance of closest approach, $d$, is limited by the constraint that gravity  must contribute at least one order of magnitude more strongly to the interaction than the dipole-dipole Casimir so that electromagnetic and other standard model interactions may be viewed as negligible contributors to the developed phase \cite{bose2017spin,van2020quantum,schut2022deco}, and it was shown in \cite{van2020quantum} that $d\sim 157\mu$m is the closest distance of approach that maintains this balance of interactions without the consideration of other techniques such as screening. Fixing $d\sim 157\mu$m, the dependence of the NDC violation may be plotted against $\beta$ (effectively $\Delta x$ in units of $d$) for a variety of values $\alpha$ parameterised by the time $t$.  Figs.\ref{fig:violation2} and \ref{fig:Fig-3-DecoViolations} reveal that one can choose {\em ranges} of times that provide large violations of the NDC (in the absence or presence of decoherence) for different superposition sizes $\Delta x$. Note, larger values of $t$ ($t >6$s) have been omitted in the figures as the NDC violations for such large $t$ suffer more strongly from decoherence effects, rendering them of limited practical use.

\end{document}